\documentclass[prx,superscriptaddress,aps,amsmath,amssymb,floatfix,reprint,raggedbottom]{revtex4-2}
\usepackage{graphicx}

\usepackage[subpreambles=true]{standalone}
\usepackage{balance}
\usepackage{xr-hyper}
\usepackage{enumitem}
\usepackage{wrapfig}

\usepackage{amsmath}
\usepackage{times}

\usepackage{dcolumn}
\usepackage{xcolor}

\usepackage{lipsum}
\usepackage{soul}
\usepackage{braket}
 
\usepackage{amsfonts}
\usepackage{multirow}

\usepackage[utf8]{inputenc}

\usepackage[ruled]{algorithm2e}

\DeclareUnicodeCharacter{2009}{\,}

\makeatletter
\newcommand*{\addFileDependency}[1]{% argument=file name and extension
  \typeout{(#1)}
  \@addtofilelist{#1}
  \IfFileExists{#1}{}{\typeout{No file #1.}}
}
\makeatother

\makeatletter 
\renewcommand{\fnum@figure}{\textbf{Fig.~\thefigure}}
\makeatother

\makeatletter
\def\bbordermatrix#1{\begingroup \m@th
  \@tempdima 4.75\p@
  \setbox\z@\vbox{%
    \def\cr{\crcr\noalign{\kern2\p@\global\let\cr\endline}}%
    \ialign{$##$\hfil\kern2\p@\kern\@tempdima&\thinspace\hfil$##$\hfil
      &&\quad\hfil$##$\hfil\crcr
      \omit\strut\hfil\crcr\noalign{\kern-\baselineskip}%
      #1\crcr\omit\strut\cr}}%
  \setbox\tw@\vbox{\unvcopy\z@\global\setbox\@ne\lastbox}%
  \setbox\tw@\hbox{\unhbox\@ne\unskip\global\setbox\@ne\lastbox}%
  \setbox\tw@\hbox{$\kern\wd\@ne\kern-\@tempdima\left[\kern-\wd\@ne
    \global\setbox\@ne\vbox{\box\@ne\kern2\p@}%
    \vcenter{\kern-\ht\@ne\unvbox\z@\kern-\baselineskip}\,\right]$}%
  \null\;\vbox{\kern\ht\@ne\box\tw@}\endgroup}
\makeatother

\emergencystretch=\maxdimen
\usepackage[compact]{titlesec}
\titlespacing{\section}{0pt}{*3}{*2}
\titlespacing{\subsection}{0pt}{*2}{*2}
\titlespacing{\subsubsection}{0pt}{*2}{*2}

\titleformat{\section}{\filcenter\normalfont\small \bfseries}{\thesection.}{1em}{\MakeUppercase}   

\newcommand{\beginsupplement}{

        \setcounter{table}{0}

        \renewcommand{\thetable}{S\arabic{table}}%

        \setcounter{figure}{0}
        \renewcommand{\thefigure}{S\arabic{figure}}%
        \setcounter{equation}{0}
        \renewcommand{\theequation}{S.\arabic{equation}}%
     }

\usepackage{booktabs}

\begin{document}

\title{Training Deep Boltzmann Networks with Sparse Ising Machines}
\par

\author{Shaila Niazi}\email{sniazi@ucsb.edu}
\affiliation{Department of Electrical and Computer Engineering, University of California, Santa Barbara, Santa Barbara, CA, 93106, USA}
\author{Navid Anjum Aadit}
\affiliation{Department of Electrical and Computer Engineering, University of California, Santa Barbara, Santa Barbara, CA, 93106, USA}
\author{Masoud Mohseni}
\affiliation{Google Quantum AI, Venice, CA 90291, USA}
\author{Shuvro Chowdhury}
\affiliation{Department of Electrical and Computer Engineering, University of California, Santa Barbara, Santa Barbara, CA, 93106, USA}
\author{Yao Qin}
\affiliation{Department of Electrical and Computer Engineering, University of California, Santa Barbara, Santa Barbara, CA, 93106, USA}
\affiliation{Google Research}
\author{Kerem Y. Camsari}\email{camsari@ece.ucsb.edu}
\affiliation{Department of Electrical and Computer Engineering, University of California, Santa Barbara, Santa Barbara, CA, 93106, USA}

\begin{abstract}
The slowing down of Moore's law has driven the development of unconventional computing paradigms, such as specialized Ising machines tailored to solve combinatorial optimization problems. In this paper, we show a new application domain for probabilistic bit (p-bit) based on Ising machines by training deep generative AI models with them. Using sparse, asynchronous, and massively parallel Ising machines we train deep Boltzmann networks in a hybrid probabilistic-classical computing setup. We use the full MNIST and Fashion MNIST (FMNIST) dataset without any downsampling and a reduced version of CIFAR-10 dataset in hardware-aware network topologies implemented in moderately sized Field Programmable Gate Arrays (FPGA). For MNIST, our machine using only  4,264 nodes (p-bits) and about 30,000 parameters achieves the same classification accuracy (90\%) as an optimized software-based restricted Boltzmann Machine (RBM) with approximately 3.25 million parameters. Similar results follow for FMNIST and CIFAR-10. Additionally, the sparse deep Boltzmann network can generate new handwritten digits and fashion products, a task the 3.25 million parameter RBM fails at despite achieving the same accuracy. Our hybrid computer takes a measured 50 to 64 billion probabilistic flips per second, which is at least an order of magnitude faster than superficially similar Graphics and Tensor Processing Unit (GPU/TPU) based implementations. The massively parallel architecture can comfortably perform the contrastive divergence algorithm (CD-$n$) with up to $n$\,=\,$10$ million sweeps per update, beyond the capabilities of existing software implementations. These results demonstrate the potential of using Ising machines for traditionally hard-to-train deep generative Boltzmann networks, with further possible improvement in nanodevice-based realizations.
\end{abstract}
\maketitle

\section{Introduction}
\label{sec:Intro}

The slowing down of Moore's Law is ushering in an exciting new era of electronics where the traditionally separate layers of the computing stack are becoming increasingly intertwined. The rise of domain-specific computing hardware and architectures is driving unconventional computing approaches. One approach that generated great excitement recently is the field of Ising machines, where special-purpose hardware is developed to solve combinatorial optimization problems \cite{Mohseni2022}. The goal of Ising machines is to improve energy efficiency, time to solution, or some other useful metric to solve optimization problems by co-designing all layers in the computing stack. 

In this paper, we draw attention to another possibility of using probabilistic Ising machines, beyond combinatorial optimization, to demonstrate their application to deep generative AI models. We focus on 
deep Boltzmann Machines (BM) that are multi-layer generalizations of the original Boltzmann Machine \cite{hinton1984boltzmann,huembeli2022physics}. Despite being powerful models, BMs fell out of favor from mainstream deep learning praxis \cite{lecun2015deep}, primarily because they are computationally hard to train with widely available hardware \cite{goodfellow2016deep}. Our goal in this paper is to illustrate how a sparse version of deep BMs can be efficiently trained using special-purpose hardware systems that provide orders of magnitude improvement over commonly used software implementations in the computationally hard probabilistic sampling task. 

With minor modifications, our core computational kernel $-$ fast probabilistic Markov Chain Monte Carlo sampling $-$ could support a large family of energy-based models, including restricted and unrestricted BMs \cite{hinton2012practical}, contrastive Hebbian learning \cite{xie2003equivalence}, Gaussian-Bernoulli BMs \cite{cho2013gaussian,liao2022gaussian}, equilibrium propagation \cite{scellier2017equilibrium}, predictive coding \cite{millidge2022backpropagation} and related algorithms. 

We design a probabilistic bit (p-bit) \cite{camsari2017stochastic} based realization of Boltzmann networks, as their lowest level realization in hardware. Using FPGAs, we physically construct a network of binary stochastic neurons (BSN) in hardware and connect them to one another in a fixed hardware topology. We also design an asynchronous architecture where p-bits (BSNs) dynamically evolve in parallel, much like an interacting collection of particles without a synchronizing global clock. Such a low-level realization of a Boltzmann network provides up to 5 orders of magnitude improvement in generating samples from the Boltzmann distribution, even in moderately sized FPGAs. An intense amount of work is currently underway to design scaled probabilistic computers out of magnetic nanodevices \cite{borders2019integer,grimaldi2022experimental,hayakawa2021nanosecond,kaiser2022hardware} which can scale probabilistic computers to much larger densities in energy-efficient implementations. Despite our FPGA-specific design in this paper, much of our results are applicable to scaled p-computers as well as other Ising machines based on many different physical realizations \cite{Mohseni2022}. Our broader goal is to help stimulate the development of physics-inspired probabilistic hardware \cite{chowdhury2023full,coles2023thermodynamic} which can lead to energy-efficient systems to reduce the rapidly growing costs of conventional deep learning based on graphics and tensor processing units (GPU/TPU) \cite{patterson2021carbon}.

\begin{figure*}[!t]
    \centering
    \includegraphics[width=1 \textwidth]{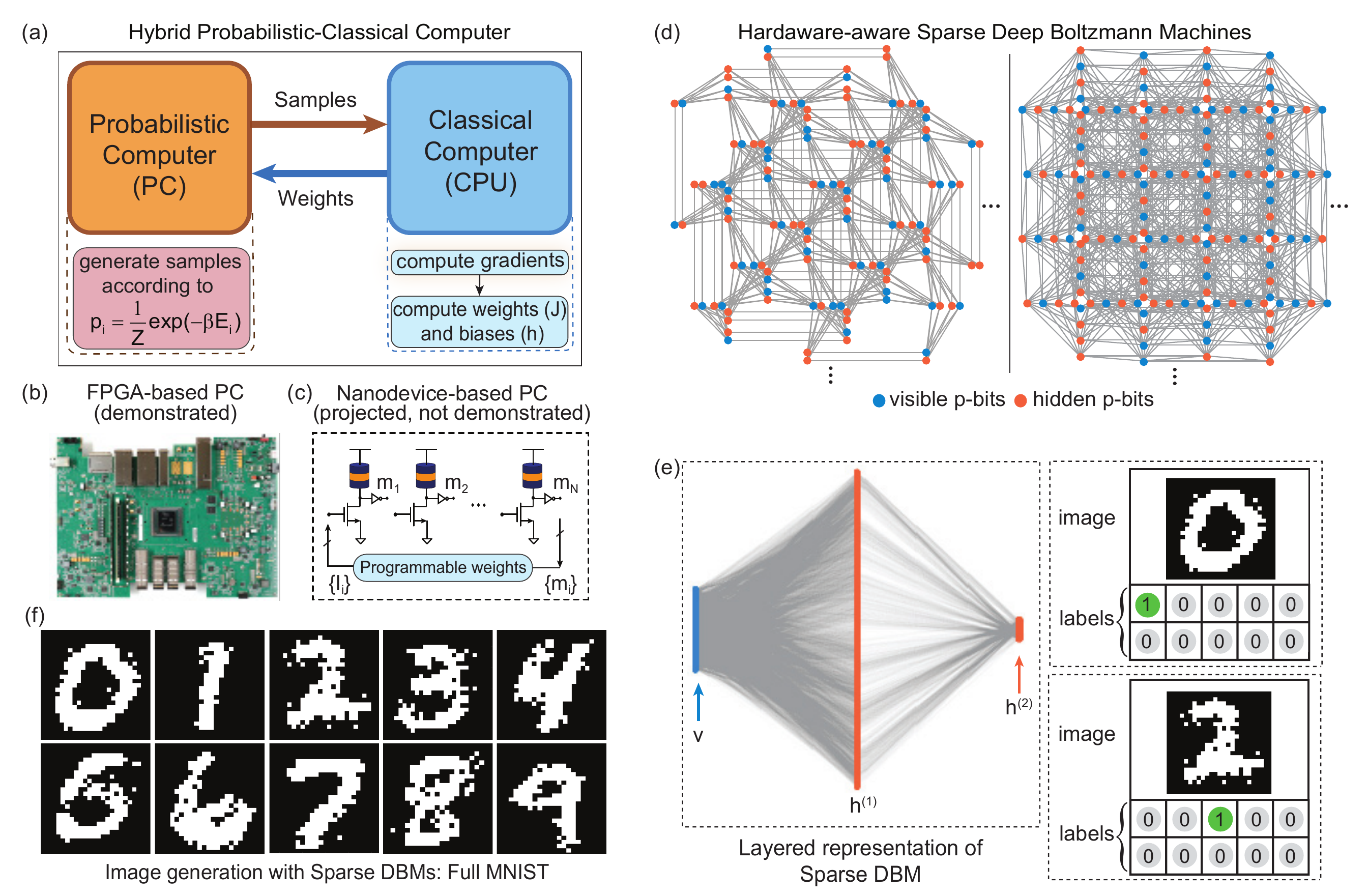}
    \vspace{-15pt}
    \caption{{\footnotesize (a) Hybrid computing scheme with probabilistic computer and classical computer implemented on a CPU. The p-computer generates samples according to the Boltzmann-Gibbs distribution and provides them to the CPU. Then CPU computes gradients, updates the weights (J) and biases (h), and sends them back to the p-computer until convergence. (b) The p-computer illustrated here is based on digital CMOS implementation (FPGA) and can have a measured sampling speed of $\approx 50\text{ to }64$ flips/ns. (c) Nanodevice-based p-computer: Various analog implementations have been proposed \cite{chowdhury2023full}. (d) Hardware-aware sparse Deep Boltzmann Machines (DBMs) are represented with visible and hidden p-bits (examples of the Pegasus \cite{dattani2019pegasus} and Zephyr graphs \cite{boothby2021zephyr} are shown). (e) The sparse DBMs shown in (d) are illustrated with two layers of hidden units (Left) where both the interlayer and intralayer (not shown) connections are allowed. (see Supplementary section~\ref{sec:actual} for a full view of the networks used in this work. The graph density and vertex degree distribution of the sparse DBMs are shown in the Supplementary Section~\ref{sec:sparsity_dbm}.) When a particular label p-bit corresponding to a digit is activated (clamping that label p-bit to 1 and clamping the rest to 0), the network evolves to an image of that digit as shown in the example (Right). (f) All 10 digits are generated with sparse DBM after training the network with the full MNIST dataset.}}
    \label{fig: overview}
    \vspace{-7pt}
\end{figure*}

\section{A Hybrid Probabilistic-Classical Computing Scheme}
\label{sec:hybrid}

The approach we take in this paper to train deep Boltzmann networks is to construct a hybrid probabilistic and classical computing setup (FIG.~\ref{fig: overview}a). The role of the classical computer is to compute gradients and the new set of weights and biases given a set of samples from the probabilistic computer.  The role of the probabilistic computer is to generate equilibrium samples for a given network (defined by the set of weights and biases) according to the Boltzmann law: 
\begin{equation}
    p_i = \frac{1}{Z} \mathrm{exp}\left(-\beta E_i\right)
   \label{eq:boltz} 
\end{equation}
where $E_i$ is the configuration dependent energy and $\beta$ is the inverse (algorithmic) temperature. In our context of probabilistic sampling, $\beta$ is typically set to 1, unlike the optimization setting where it is gradually increased to find the configuration with minimum energy. In general, the configuration-dependent energy can be expressed as a $k$-local Hamiltonian \cite{sejnowski1986higher}, in this paper, we focus on the 2-local energy that is given by: 
\begin{equation}
     E = - \left(\sum_{i<j} J_{ij} m_i m_j + \sum h_i m_i \right) \vspace{-4pt}
     \label{eq:en}
\end{equation}
where $J_{i j}$ and $h_i$ represent the network topology and $m_i$ represents the bipolar state of nodes that are either $+1$ or $-1$. The probabilistic computer we design approximates the Boltzmann law by the following dynamical equations, where the effective field $I_i$ and the activation of $m_i$ are given by \cite{camsari2017stochastic,faria2021hardware}: 
\begin{equation}
    I_i (t+\Delta t) = \sum J_{ij} m_j (t) + h_i 
    \label{eq:synapse}
\end{equation}
and the activation of a p-bit is given: 
\begin{equation}
 m_i (t) = \mathrm{sgn}(\mathrm{tanh}[\beta I_i(t)]- \mathrm{rand}_{\text{U},[-1,1]})
 \label{eq:pbit}
\end{equation}
The iterated evolution of Eq.~(\ref{eq:synapse}) and Eq.~(\ref{eq:pbit}) with a predefined (or random) update order generates samples approximating the Boltzmann law defined by Eq.~(\ref{eq:boltz}) \cite{aarts1989simulated}. Note that in the rest of this paper $\beta$ is always set to 1, except in image generation experiments where we anneal the network. 

An important requirement to reach the Boltzmann equilibrium is that connected p-bits are updated serially (re-computing Eq.~\eqref{eq:synapse} every time) rather than in parallel \cite{Aadit2022a} so that each p-bit updates with the most up-to-date information. This iterative process is called Gibbs Sampling \cite{koller2009probabilistic} and is a fundamental Markov Chain Monte Carlo (MCMC) algorithm used in many machine learning applications \cite{andrieu2003introduction}. The physical implementation of Eq.~(\ref{eq:synapse}) and Eq.~(\ref{eq:pbit}) to perform MCMC introduces several challenges. The primary difficulty is the serial updating requirement of connected p-bits, prohibiting the parallelization of updates in dense networks. The second difficulty is to ensure p-bits receive all the latest information from their neighbors before updating, otherwise, the network does not sample from the true Boltzmann distribution \cite{pervaiz2017hardware}. 

\section{Hardware-aware Sparse Networks}
\label{sec:hw}

Both of these difficulties are more easily addressed in sparse networks. Sparsity limits the number of neighbors between p-bits allowing parallel and asynchronous updates on unconnected p-bits. Indeed, we show that as long as the chosen network topology is sparse, a massively parallel architecture where the frequency of probabilistic samples that linearly depends on the number of nodes in the network can be constructed \cite{Aadit2022a} (see Section~\ref{sec:architecture} for details). Our present FPGA implementation of the probabilistic computer can take up to $50$ to $64$ flips per nanosecond (flips/ns) and projections indicate stochastic magnetic tunnel junction-based (sMTJ) implementations can take this number to about a million flips/ns or more (FIG.~\ref{fig: overview}b,c) \cite{sutton2020autonomous,camsari2017implementing}. These projections have not been realized, however, nanosecond fluctuations with sMTJs \cite{hayakawa2021nanosecond,safranski2021demonstration} have been demonstrated.  Given the gigabit densities in MTJ-based memory \cite{lee20191gbit}, large-scale integration of p-computers remains plausible,  with MTJ-based prototypes demonstrating architectures of the type we discuss here \cite{borders2019integer,kaiser2022hardware,grimaldi2022experimental}.

In this paper, we adopt the Pegasus \cite{dattani2019pegasus} and the Zepyhr \cite{boothby2021zephyr} topologies developed by D-Wave's quantum annealers to train hardware-aware sparse deep BMs (FIG.~\ref{fig: overview}d). Even though our approach is applicable to any sparse graph (regular and irregular), we focus on such hardware-aware networks with limited connectivity where maximum degrees range between 15 and 20. Our choice of sparse models is motivated by scaled but connectivity-limited networks such as the human brain and advanced microprocessors. 

Despite the common use of full connectivity in BM-based networks where inter-layer connections are typically fully connected \cite{salakhutdinov2009deep}, both advanced microprocessors with networks of billion transistors and the human brain exhibit a large degree of sparsity \cite{bassett2006small}. In fact, most hardware implementations of RBMs \cite{bojnordi2016memristive,tsai201741,kim2009highly} suffer from scaling problems due to large fan-outs, requiring off-chip memory access or distributed computation in multiple chips \cite{kim2009highly}. On the other hand, sparse connectivity in hardware neural networks often exhibits energy and area advantages \cite{ardakani2016sparsely}.

FIG.~\ref{fig: overview}e shows a typical sparse DBM that we use in this paper with 2-layers of hidden bits. This graph is obtained by randomly assigning visible and hidden bits in the Pegasus (or Zephyr) graphs of various sizes. Unlike standard deep BMs \cite{salakhutdinov2010efficient,goodfellow2013joint}, sparse DBMs do not have fully-connected interlayer connections. On the other hand, they do allow connections between nodes in a given layer, increasing the representative capability of the network. 
In Section~\ref{sec:random}, we systematically study the effect of distributing visible/hidden nodes in such sparse networks, which introduces new challenges that do not exist in fully connected networks. 

Unlike standard deep BM training where training is typically done layer-by-layer \cite{hinton2002training}, in sparse DBMs, we tackle the training directly on the full network, by relying on our massively parallel architecture and the efficient mixing of sparse graphs.

As we discuss in Section~\ref{sec:train}, we reach about 90\% classification accuracy in 100 epochs with the full MNIST dataset without any downsampling, coarse-graining, or the use of much simpler datasets, typically performed in alternative hardware-based approaches \cite{adachi2015application,manukian2019accelerating,dixit2021training,bohm2022noise}. To support our conclusions, we also train the harder fashion MNIST dataset and a reduced version of CIFAR-10, in the Supplementary Section~\ref{sec:fashionMnist}-\ref{sec:cifar100}.

Moreover, unlike RBMs, the sparse DBM learns the images well enough that for any given label, it can generate a new handwritten digit (or an FMNIST sample) as shown in FIG.~\ref{fig: overview}f, when a single one-hot encoded output p-bit is clamped to a given digit. 
 Image generation is an important feature of physics-inspired algorithms such as diffusion models \cite{sohl2015deep}, and the fact that RBMs fail at this task even when they have 100$\times$ more parameters is surprising (both in MNIST and FMNIST), stressing the potential of sparse DBM models, as we discuss further in Section~\ref{sec:ImageSynth}.

\section{Training Sparse DBMs with Sparse Ising Machines}
\label{sec:algorithm}

\SetKwComment{Comment}{\normalfont $\triangleright$ }{}
\SetKwInOut{Input}{Input}
\SetKwInOut{Output}{Output}
\SetKwInput{KwSampler}{Sampler}
\begin{algorithm}
\caption{Training sparse DBMs}
\label{alg:alg2}
\Input{number of samples $N$,  batch size $B$, number of batches $N_{B}$, epochs $N_{\text{L}}$, learning rate $\varepsilon$}
\KwSampler{Classic Gibbs CPU, Graph-colored Gibbs CPU, Graph-colored Gibbs FPGA}
\Output{trained weights $J_{\text{out}}$ and biases $h_{\text{out}}$}
$J_{\text{out}} \gets \mathcal{N}(0,0.01)$\;
$h_{\text{out,hidden}} \gets 0$\;
$h_{\text{out,visible}} \gets \log{(p_i/(1-p_i))}$\;
\For{$i\gets 1$ \KwTo $N_{\text{L}}$}{
    \For{$j \gets 1 $ \KwTo $N_{\text{B}}$}{
        $J_{\text{Sampler}} \gets J_{\text{out}}$\;
        \tcc{positive phase}
        \For{$k \gets 1$ \KwTo $B$}{
            $h_{\text{$B$}} \gets \text{ clamping to batch images}$\;
            $h_{\text{Sampler}} \gets h_{\text{$B$}}+h_{\text{out}}$\;
            $\{m\} \gets \text{Sampler}(N)$ \Comment*[r]{p-computer}\ 
            }
            $\langle m_im_j\rangle_{\text{data}} = \{m\}\{m\}^{\text{T}}/(N\times B)$\ \Comment*[r]{CPU}\
        \tcc{negative phase}
        $h_{\text{Sampler}} \gets h_{\text{out}}$\;
        $\{m\} \gets \text{Sampler}(N\times B)$ \Comment*[r]{p-computer}\ 
        $\langle m_im_j\rangle_{\text{model}} = \{m\}\{m\}^{\text{T}}/(N\times B)$\ \Comment*[r]{CPU}\ 
        \tcc{update weights and biases} 
        $J_{\text{out}} \gets J_{\text{out}} + \Delta J_{ij}$\ \Comment*[r]{CPU}\ 
        $h_{\text{out}} \gets h_{\text{out}} +\Delta h_i$\ \Comment*[r]{CPU}\ 
    }
}
\end{algorithm}

\begin{figure*}[t!]
    \centering
    \includegraphics[width=1\linewidth]{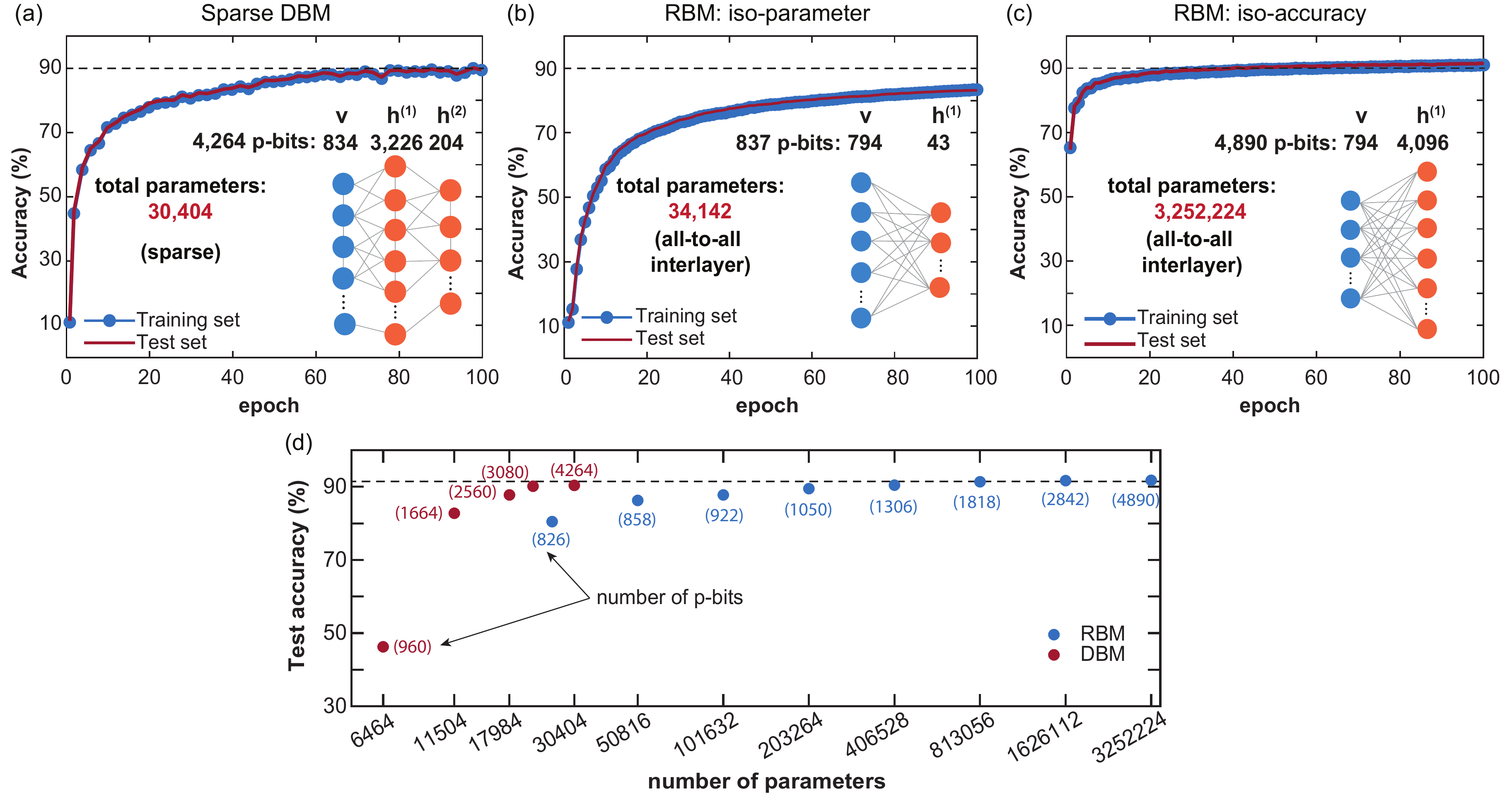}
    \vspace{-15pt} 
    \caption{{\footnotesize (a) MNIST accuracy vs training epochs: with sparse DBM, 90\% accuracy is achieved in 100 epochs. Full MNIST (60,000 images) is trained on sparse DBM (Pegasus 4,264 p-bits) with CD-$10^{5}$, batch size = 50, learning rate = 0.003, momentum = 0.6 and epoch = 100 where the total number of parameters is 30,404. Each epoch is defined as the network seeing the entire 60,000 images with 1,200 weight updates. Test accuracy shows the accuracy of all the 10,000 images from the MNIST test set and the training accuracy represents the accuracy of 10,000 images that are randomly chosen from the training dataset. (b) MNIST accuracy with Restricted Boltzmann Machine (RBM) using 43 hidden units and CD-1 (CPU implementation) where the total number of parameters is 34,142. The accuracy of this RBM is less than 90\% but sparse DBM can reach 90\% with approximately the same number of parameters. (c) MNIST accuracy of RBM with 4,096 hidden units. Here the total number of parameters is 3,252,224 and the accuracy is 90\% in 100 epochs which can be achieved using sparse DBM with around $100\times$ fewer parameters. (d) Test accuracy of MNIST as a function of the number of parameters with sparse DBMs (Pegasus) and RBMs. We trained full MNIST with 5 different sizes of Pegasus graphs for 100 epochs using the same set of hyperparameters and collected the test accuracy of the whole test set. When the number of parameters is only 6,464 with the smaller Pegasus (960 p-bits), test accuracy could not reach beyond 50\%. On larger graphs with increased parameters, accuracy starts to increase and $\approx$ 90$\%$ accuracy is achieved with the largest Pegasus (4264 p-bits) that fits into our FPGA. RBM reached 90\% accuracy with around 200,000 parameters but the increased number of parameters (up to 3.25 million) could not help go beyond $\approx 92\%$ accuracy.}} 
    \label{fig:accuracy}
    \vspace{-10pt}
\end{figure*}

As our network model, we use the Pegasus \cite{dattani2019pegasus} and Zephyr \cite{boothby2021zephyr} graphs at different sizes as a fixed network. 
Boltzmann networks are typically used for unsupervised learning without any explicit labels. To use deep BMs for classification, we follow a similar approach to \cite{larochelle2008classification} where we create additional visible bits, calling them ``label bits''. We use one-hot encoding to specify 10 digits with 10 label bits, such that each image in the MNIST data set is paired with these label bits (FIG.~\ref{fig: overview}e). We then use our fast Gibbs sampler (probabilistic computer) to perform the contrastive divergence (CD) algorithm \cite{hinton2002training,larochelle2007empirical} that minimizes the KL divergence between the data and the model distributions. An equivalent formulation from a maximum likelihood estimation viewpoint \cite{koller2009probabilistic,pmlr-vR5-carreira-perpinan05a} can also be used to obtain the following learning rule (see Supplementary Section~\ref{sec:derivation}),
\begin{eqnarray}
\Delta J_{ij} &=& \varepsilon\bigg(\langle m_im_j\rangle_\text{{data}}-\langle m_im_j\rangle_\text{{model}}\bigg) \label{eq:del_J}\\
\Delta h_{i} &=& \varepsilon\bigg(\langle m_i\rangle_\text{{data}}-\langle m_i\rangle_\text{{model}}\bigg)
   \label{eq:del_h}
\end{eqnarray}
where $\Delta J_{ij}$ and $\Delta h_{i}$ represent the weight and bias updates per iteration and the terms in the parentheses represent the negative gradient of the KL divergence between data and the model distributions. $\varepsilon$ is the learning rate, $\langle m_im_j\rangle_\text{{data}}$ and $\langle m_im_j\rangle_\text{{model}}$ are the average correlation between p-bits $i$ and $j$ in the ``positive'' (data) and ``negative'' (model) phases, respectively. During the positive phase of sampling, the p-computer clamps the visible p-bits to the corresponding training image one after the other, taking $N$ sweeps for each image for a total of $N \times B$ sweeps where $B$ is the batch size. Using these sweeps, the CPU then computes the data correlations $\langle m_im_j\rangle_{\text{data}}$. In the negative phase, the p-computer is allowed to run freely without any clamping, and the CPU computes the model correlations $\langle m_im_j\rangle_{\text{model}}$ by taking $N\times B$ sweeps. Then the connection weights are updated according to Eq.~\eqref{eq:del_J} and Eq.~\eqref{eq:del_h}. In actual training, we also use a momentum modification to Eqs.~(\ref{eq:del_J},\ref{eq:del_h}) (see Supplementary Section~\ref{sec:momentum}).  A pseudocode of the algorithm is presented in Algorithm~\ref{alg:alg2}.

For the sparse DBMs we consider in this work, establishing correlations between the data requires executing Gibbs sampling even for the positive phase, which is obtained in a single inference step in RBMs. Our machine can be configured to implement the persistent contrastive divergence (PCD) \cite{tieleman2008training,hinton2012practical,fischer2014training} algorithm. PCD maintains a long-running Markov chain such that small changes in weights do not take the equilibrium state of the new network far from the old one. We discuss the possible benefits of PCD vs CD in the context of our results in Section~\ref{sec:equilibrium}.

\section{Results on the Full MNIST dataset}
\label{sec:train}

The dataset that we used for training sparse DBMs is the full MNIST handwritten digit dataset \cite{lecun1998mnist,deng2012mnist} without any reduction or downsampling. We show results on FMNIST and CIFAR-10 in the Supplementary Section~\ref{sec:fashionMnist}-\ref{sec:cifar100}. MNIST consists of 60,000 training images and 10,000 test images with $28\times28$ pixels having digits from 0 to 9 and we use black/white images by rounding up the pixel intensities. We set the initial values of weights and biases according to the recipe that Hinton suggested for RBMs in \cite{hinton2012practical}. The weights are initialized to small random values chosen from a zero-mean Gaussian with a standard deviation of 0.01 for every p-bit. The initial values of hidden biases are set to zero and visible biases are set to log[$p_i/(1-p_i)$] where $p_i$ is the proportion of training vectors in which unit $i$ is on. The values of hyperparameters used while training are  $\beta$ = 1, learning rate $\varepsilon$ = 0.003, and momentum $\alpha$ = 0.6.
 \begin{figure*}[!t]
    \centering
    \includegraphics[width=.85\linewidth]{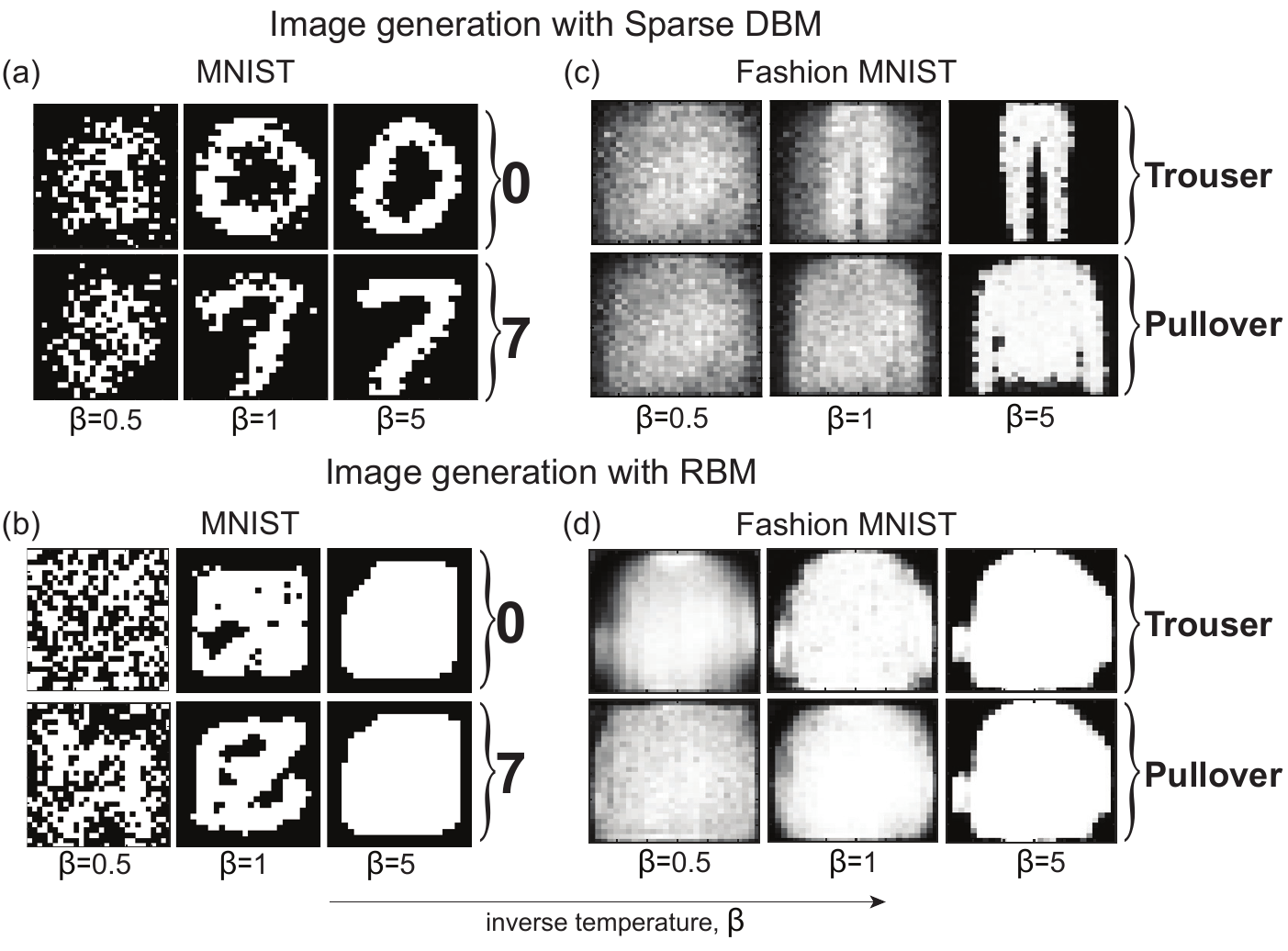}
    \vspace{-10pt}
    \caption{{\footnotesize (a) Images generated with sparse DBM by annealing the network from  $\beta$\,=\,0 to  $\beta$\,=\,5 with 0.125 steps after training the full MNIST dataset. The labels for a particular digit are clamped to show how the visible p-bits evolve to that specific image. Examples of digits `0' and `7' are shown here. (b) The same procedure for image generation is applied to the RBM network (with 4,096 hidden units) that achieves 90\% test accuracy. Using the same annealing schedule, RBM does not produce the correct digits, unlike the sparse DBM. (c) Generated images of fashion products (e.g. `Trouser' and `Pullover') with sparse DBM by annealing the network from  $\beta$\,=\,0 to  $\beta$\,=\,5 with 0.125 steps after training full Fashion MNIST. (d) RBM with 4096 hidden units can not generate the correct images according to the labels despite achieving around 83\% test accuracy}.}
    \label{fig:ImSynth}
    \vspace{-8pt}
\end{figure*}

The sparse DBM network used here (the largest size Pegasus that we can fit into our FPGA) consists of 834 visible p-bits ($834 = 28 \times 28 + 10 \times 5$; we used 5 sets of labels each containing 10 p-bits) and 3,430 hidden p-bits arranged in 2-layers as shown in the inset of FIG.~\ref{fig:accuracy}a. Then we randomly distribute the visible and hidden units on the sparse DBMs to ensure the label indices are delocalized (see Section~\ref{sec:random} for details of this process and the original network in Supplementary FIG.~\ref{fig:P14}). 

To train the network efficiently, we divide the training set into 1,200 mini-batches having 50 images in each batch. The weights are updated after each mini-batch following the CD algorithm. We train MNIST for 100 epochs with CD-$10^{5}$, where $10^5$ sweeps of the entire network are taken in the negative phase ($N\times B$). The weight precision in the FPGA is 10 bits (1 sign, 6 integer, and 3 fraction bits, i.e., s\{6\}\{3\}) while the CPU uses double-precision floating-point with 64 bits, to compute the gradients. Before the new weights are loaded into the FPGA, however, they are reduced to s\{6\}\{3\} to fit into the FPGA. A systematic study of the effect of weight precision is shown in Supplementary Section~\ref{sec:bit_precision} along with image completion experiments. In short, we do not observe any significant differences at higher precision in the FPGA, indicating that the 10-bit weight precision is adequate. 

During inference, the 784 p-bits that correspond to the pixels are clamped to the test data and the label p-bits fluctuate freely. To test classification accuracy, we use $10^5$ sweeps and perform a softmax classification scheme as follows: as we have 50 label p-bits for 5 sets of labels, by time-averaging the corresponding label bits we finally have the 10 labels for 10 digits. The p-bit with the highest probability of being `1' is used for the classified digit. For comparison, we also train an optimized RBM model using CD-1 in the CPU. The label, testing, and training details of RBMs are very similar to those of sparse DBMs.

FIG.~\ref{fig:accuracy} shows our main results. We see that the sparse DBM architecture in the Pegasus graph with 4,264 p-bits reaches about 90\% accuracy in 100 epochs (see Supplementary Section~\ref{sec:train_100im} where the training accuracy can reach 100\% for MNIST/100 images).  To compare the sparse DBM architecture with a standard RBM, we perform two tests, one at ``iso-parameter'' and the other at ``iso-accuracy''.  The iso-parameter test uses an RBM with about the same number of parameters (with an all-to-all interlayer connection). This RBM falls short of reaching 90\% in this setting. Then, we choose an RBM with 100$\times$ more parameters and observe that the results saturate at about 90\% accuracy. We also note that increasing CD-1 to CD-$n$ ($n$ up to 100) does not result in an appreciable difference in accuracy while making the training computationally much harder.

Detailed testing in both models (sparse DBM and RBM) indicates that marginal improvements are possible with more training epochs, however, both models show similar asymptotic behavior in 100 epochs, this is why we stop training around 100 epochs (FIG.~\ref{fig:accuracy}d shows experiments at various network sizes).  Note that this is still a computationally intense process where 60,000 images are shown to the network for a total of 6,000,000 times and the weights are updated a total of 100$\times N_B$ = 120,000 times since $N_B$\,=\,1200.

To investigate the effect of total parameters of sparse DBMs on the accuracy, we used five Pegasus graphs of different sizes to train MNIST using our massively parallel architecture. These include 960, 1,664, 2,560, 3,080, and 4,264 p-bit graphs with a varying number of parameters from $\approx 6,000$ to $\approx 30,000$ as shown in FIG.~\ref{fig:accuracy}d. We trained full MNIST on each of these five sparse DBMs with CD-{$10^5$} using the same hyperparameters and reported the classification accuracy for the entire test set.  Similarly, we also trained eight different RBMs with full MNIST for 100 epochs to compare their accuracy with the number of parameters (FIG.~\ref{fig:accuracy}d). Increasing the number of parameters to millions could not increase the test accuracy significantly whereas 90\% accuracy is achieved with around 200,000 parameters.

Based on these experimental results, we arrive at the following two important conclusions: First, the sparse DBM architecture despite having a much smaller degree of connections between its layers (limited to a graph degree of $\approx$15 to 20) matches the classification accuracy of a fully-connected RBM. Second, the sparse DBM requires far fewer parameters (about 30,000) to reach 90\% accuracy in the MNIST dataset. Both of these indicate the potential of sparse DBMs which can be directly tackled by the orders of magnitude acceleration obtained in the hardware. We show in the Supplementary Section~\ref{sec:fashionMnist}-\ref{sec:cifar100} that similar results with the same order of magnitude differences between sparse DBMs and RBMs hold for the full FMNIST and a reduced version of the CIFAR-10 dataset.

Compared to more powerful standard DNN algorithms such as CNNs,  sparse DBMs do not reach state-of-the-art classification accuracy in MNIST at these modest network sizes and depths. Further improvements should be possible by algorithmic techniques and at larger sizes as discussed in Ref's \cite{larochelle2012learning,tieleman2008training}. Surprisingly, however, in a head-to-head comparison using the same contrastive divergence algorithm, the sparse DBM architecture matches the performance of highly optimized RBMs, despite the severely limited connectivity. More detailed comparisons may reveal the true potential of hardware-aware sparse DMBs which can be implemented on Ising Machines. It is important to note that the generative nature of BMs allows applications beyond classification, such as representing many-body quantum wavefunctions \cite{dawid2023introduction,carleo2017solving}. 

\section{Image generation}
\label{sec:ImageSynth}

Given their generative nature, a natural question to ask is whether the sparse DBM and RBM can generate new images when they are run in the ``invertible mode''. This is similar to the image generation from ``noise'' discussed in diffusion models \cite{sohl2015deep}. We test this idea post-training by clamping the label bits to a given digit and annealing the network using the inverse temperature, $\beta$.

Here we present an example of image synthesis with sparse DBMs with $\approx$\,30,000 parameters and the optimized RBM with $\approx$\,3.25 million parameters (FIG.~\ref{fig:ImSynth}a-b for MNIST and FIG.~\ref{fig:ImSynth}c-d for Fashion MNIST). For this process, we clamp the label bits for digits `0' or `7' in the case of MNIST while all other p-bits run freely. Using the final weights and biases and by annealing the network slowly from $\beta$\,=\,$0$ to $\beta$\,=\,$5$ with a $0.125$ increment and we check the 784 image p-bits at various time steps. At lower values of  $\beta$ when the system is in a high-temperature state, the model is sampling from noise (first column of FIG.~\ref{fig:ImSynth}a). With increasing $\beta$ values, digits gradually become recognizable, and at final $\beta$\,=\,$5$ we see clear images of a `0' or `7' (leftmost column of FIG.~\ref{fig:ImSynth}a). This example is demonstrated with the Pegasus graph using about 30,000 parameters from FIG.~\ref{fig:accuracy}, similar results are obtained with the Zephyr graph as we discuss in Supplementary Section~\ref{sec:ImSynth_zephyr}. Using the same approach of image generation, the fashion products `Trouser' or `Pullover' from FMNIST are generated as shown in FIG.~\ref{fig:ImSynth}c (details in Supplementary Section~\ref{sec:fashionMnist}).

In contrast, the images generated with the RBM (4,096 hidden units) are not recognizable even after careful annealing (FIG.~\ref{fig:ImSynth}b-d), despite multiple experiments with different trials. Similarly, the annealing schedule for RBM varies from $\beta$\,=\,$0$ to $\beta$\,=\,$5$. To test whether RBMs can generate images with better gradient calculations, we also trained an RBM with 4,096 hidden p-bits with CD-$10^{2}$ but this did not lead to any success in image generation. Interestingly, however, the RBM with 4,096 hidden p-bits and CD-$10^{2}$ can accomplish a simpler ``image completion'' task when it is presented half of a given digit (see Supplementary Section~\ref{sec:im_compltn_dbm_rbm}).  

These results seem to be in keeping with the idea that in freely ``dreaming'' or image-completing networks, it may be possible to generate images with RBMs \cite{hu2016deep}. In our experiments, the image generation task is forced by clamping only the label bits, without giving the network any other lead. In this stricter sense, the failure of the RBM to generate images is consistent with the general understanding on the subject \cite{goodfellow2020generative,sleeman2020hybrid}. We believe that accelerating the Gibbs sampling by orders of magnitude can enable the training of even deeper, previously untrainable deep BMs. The potential of physics-inspired RBMs for image generation is also seen in the recent interest in Gaussian-Bernoulli (GRBMs) \cite{liao2022gaussian}, whose sparse and deep variants could be even more powerful.

\section{Mixing times}
\label{sec:equilibrium}

One of the key difficulties that are often cited in the training of Boltzmann networks is the computational intractability of the partition function, $Z$, that appears in the Boltzmann law, Eq.~(\ref{eq:boltz}). Formally, what is required to ensure an exact calculation of the gradient is that the calculation of correlations $\langle m_i m_j \rangle$ and averages $\langle m_i \rangle$ come from the equilibrium states of a given network defined by $J_{ij}$ and $h_i$. The time it takes for an undirected network to reach equilibrium is defined as the ``mixing time''. A formal analysis of how long it takes for a given graph to mix can be extremely difficult to calculate and is unknown for all but the simplest, most regular networks \cite{levin2017markov}. Here, we empirically study the mixing time of the Pegasus graph that we used in generating our main results in FIG.~\ref{fig:accuracy}. The fact that there is no a priori method to determine mixing times, a lot of hyperparameter optimization might be necessary to squeeze the maximum results out of these networks (see, for example, \cite{hinton2012practical}).

\begin{figure}[!t]
    \centering
    \includegraphics[width=0.85\columnwidth]{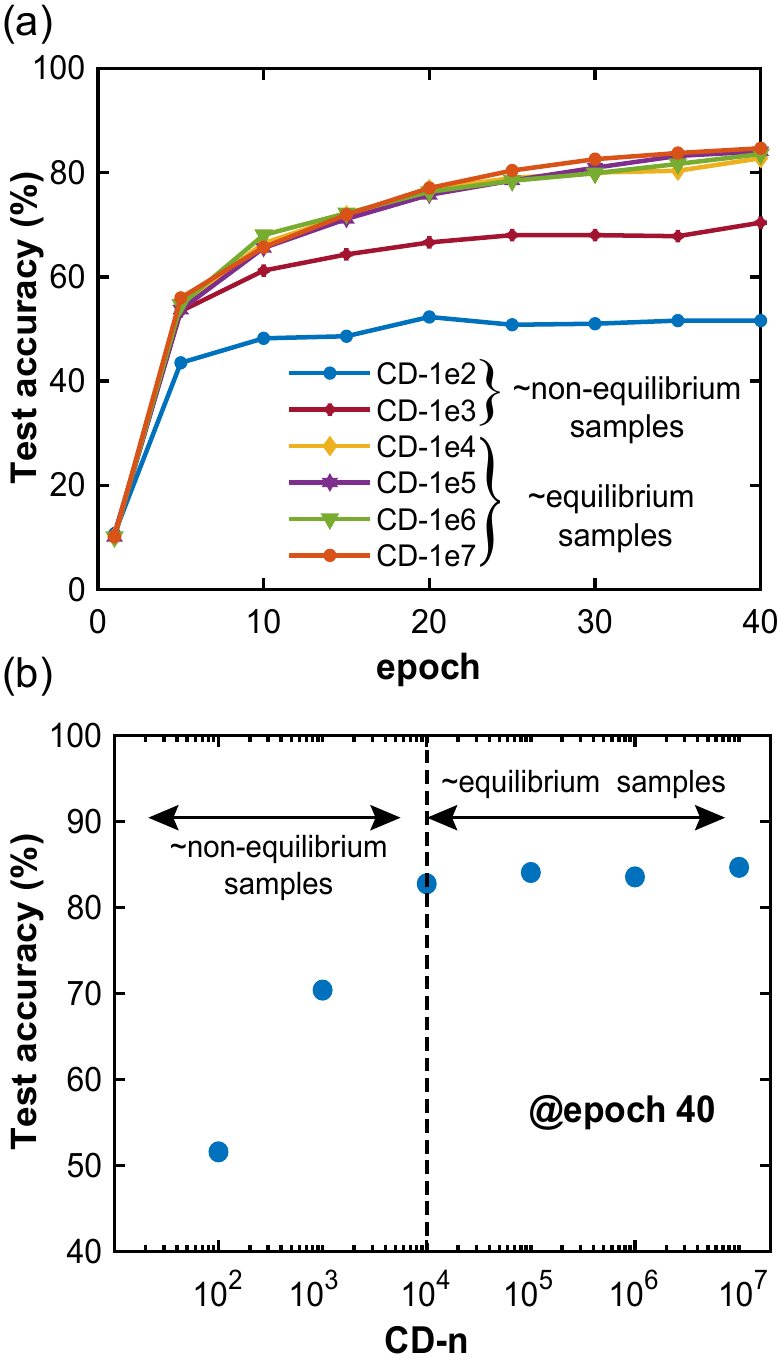}
    \vspace{-10pt} 
    \caption{{\footnotesize (a) Test accuracy after training full MNIST (up to only 40 epochs for computational simplicity) with different numbers of sweeps per iteration is shown. For our sparse graph, to mix the Markov chain properly we need a minimum CD-$10^{4}$. Reducing the number of sweeps to $10^{3}$ or $10^{2}$ degrades the quality of mixing the chain significantly. (b) Test accuracy as a function of CD-n at epoch 40 showing the equilibrium and non-equilibrium samples.}}     \label{fig:real_sweep}
    \vspace{-10pt}
\end{figure}

\begin{figure*}[!t]
    \centering
    \includegraphics[width =.99\textwidth ]{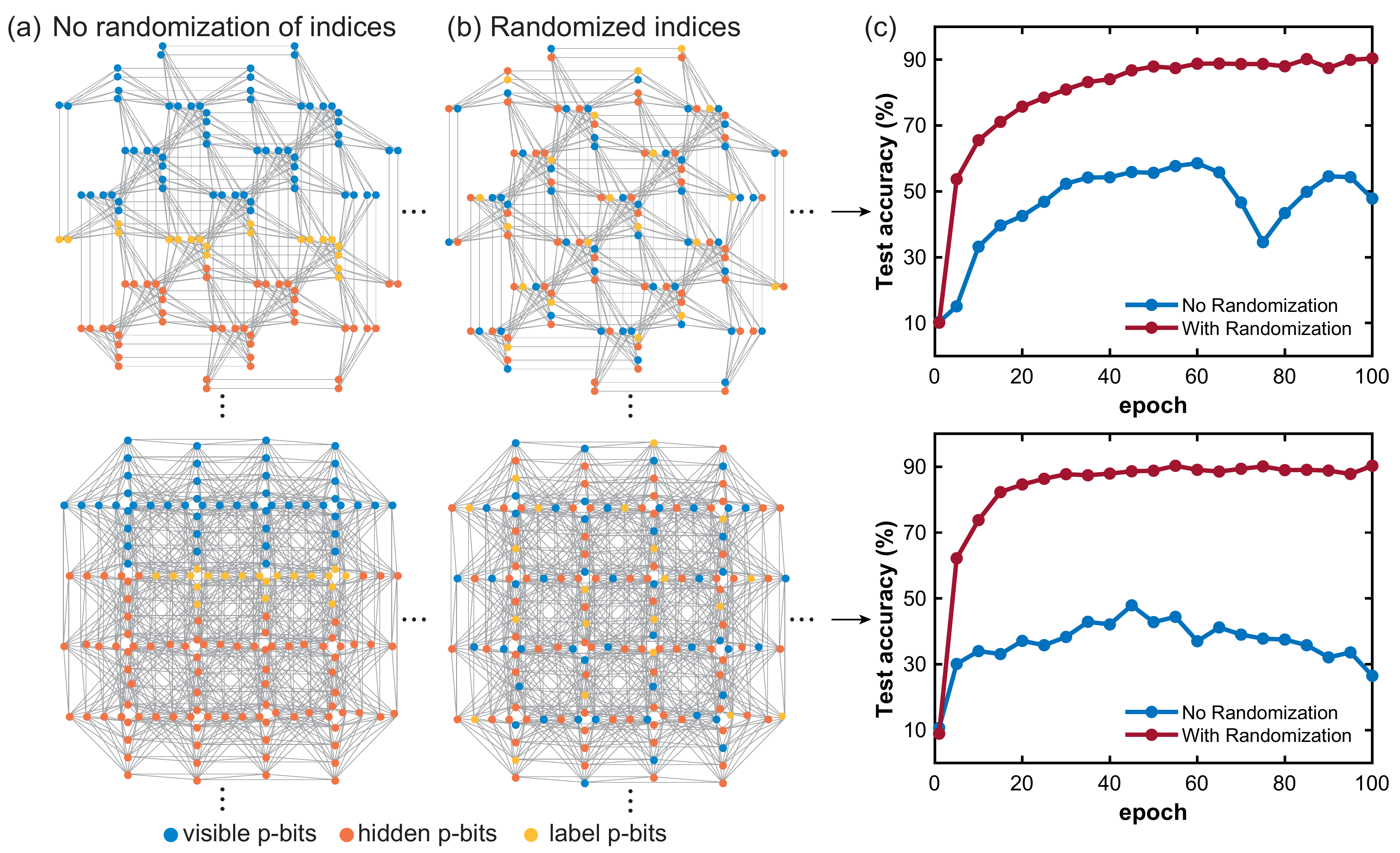}
    \caption{{\footnotesize (a) The sparse DBMs (Pegasus and Zephyr) where all the p-bits are distributed in a serial manner such as 1 to 784 are the visible p-bits, 785 to 834 are the label p-bits (50 bits for 5 sets of labels), and the rest are hidden p-bits. (b) The sparse DBMs with randomized indices are shown here. (c) Test accuracy of full MNIST as a function of training epochs for two different sparse DBMs. In both cases, training the sparse DBMs with the serial distribution (no randomization) of indices could not achieve an accuracy of more than 50\%. In contrast, randomization of indices helps the network to reach ~90\% accuracy.}}
\label{fig:rand_label}
\vspace{-8pt}
\end{figure*}

In FIG.~\ref{fig:real_sweep}, we observe that the test set accuracy of our network increases significantly if the probabilistic sampler takes $10^4$ or more sweeps per weight update. Above this value, there seem to be diminishing returns in improving the accuracy. This suggests that taking more sweeps does not improve the estimation of the averages and correlations because these samples are already in equilibrium and $\approx$\,$10^4$ sweeps at this size (with 4,264 p-bits) of the Pegasus graph can be empirically defined as the mixing time of the network (Supplementary Section~\ref{sec:mixing_P3080} shows the mixing time study of different size Pegasus). As mentioned earlier, our probabilistic computer could be modified to perform persistent CD algorithm (PCD) \cite{tieleman2008training}. Beyond CD-$10^4$, this may have diminishing returns since the chain mixes and starts sampling from the equilibrium distribution, even if it starts from a random state, as shown in FIG.~\ref{fig:real_sweep}.

The reason for the saturating classification accuracy of sparse DBMs at around 90\%  is likely that the network is not deep or wide enough and not because of the intractability of the algorithm. In fact, considering our hardware architecture FPGA is able to take $\approx$ 64 billion samples per second, and obtaining $10^5$ sweeps from our machine can be done in mere milliseconds (Table~\ref{tab:fps} shows comparisons of sampling rates between standard CPUs and our graph colored (GC) architecture, where our probabilistic computer (GC-FPGA) demonstrates $\approx$ 4 to 6 orders of magnitude improvement over the optimized and standard CPU implementations of Gibbs sampling, respectively). In Supplementary Section~\ref{sec:gpu}, we show how the performance reported in Table~\ref{tab:fps} fares against superficially similar Ising solvers in highly optimized GPU and TPU implementations.

These results suggest that our machine can be used to sample much more ``difficult'' networks that require many more samples to mix, enabling the training of previously untrainable networks with potentially richer representation. 

\begin{table}[!b]
    \centering
    \caption{{\footnotesize Comparison of the FPGA-based MCMC sampler with standard CPU and graph-colored CPU implementations. All data points are measured, as discussed in the Methods.}}
    \vspace{7.5pt}
    \begin{tabular}{@{}lcccc@{}}
        \toprule
        {\bf Sampling method} & {\bf topology} &{\bf  size} & {\bf max. degree} & {\bf flips/ns} \\ \midrule

         Standard Gibbs (CPU) & Pegasus & 4,264 & 15 & $2.18\times10^{-5}$ \\
         GC Gibbs (CPU) & Pegasus & 4,264 &  15 & $8.50\times10^{-3}$\\
         GC Gibbs (FPGA) &  Pegasus & 4,264 & 15 & $6.40 \times10^{1}$ \\
         \midrule
         Standard Gibbs (CPU) & Zephyr & 3,360 &  20 & $2.67\times10^{-5}$ \\
         GC Gibbs (CPU) & Zephyr & 3,360 & 20 & $4.40\times10^{-3}$\\
         GC Gibbs (FPGA) &   Zephyr & 3,360 & 20 & $5.04\times10^{1}$\\
         \bottomrule
        
    \end{tabular}
    \label{tab:fps}
\end{table}

It is important to note that in case of the contrastive divergence algorithm where the goal is to estimate model correlations and averages, one does not need to compute the effect of all samples. After the network reaches equilibrium, a small number of samples may be used to estimate the averages and correlations (with $\mathcal O(1/\sqrt{N})$ complexity). This significantly eases practical read-out constraints of a fast probabilistic sampler (See Supplementary Section~\ref{sec:read_out} for our detailed read-out architecture). 

\section{Randomization of indices}
\label{sec:random}
A very important point that arises in training Boltzmann network models on a given sparse network is the notion of graph distance between visible, hidden, and label nodes. Typically, if the layers are fully connected, the graph distance between any given two nodes is a constant. On the other hand, when training sparse DBMs, the placement of visible, hidden, and label p-bits plays a crucial role. FIG.~\ref{fig:rand_label} shows the comparison of how the indexing of p-bits affects the classification accuracy in the Pegasus and Zephyr graphs. We observe that if the hidden, visible, and label bits are clustered and too close, the classification accuracy suffers greatly. This is likely because the correlation between the label bits and the visible bits gets weaker if their graph distance is too large. On the other hand, randomizing the indices seems to solve this problem, repeatable in completely different but sparse graphs. We performed further experiments with different random indices and essentially observed the same behavior. For example,  FIG.~\ref{fig:accuracy}d shows a monotonically increasing accuracy with different sizes of sparse DBMs (Pegasus) even though each graph has a different randomized index set. 

\begin{figure*}[!t]
    \centering 
\includegraphics[width=0.99\textwidth]{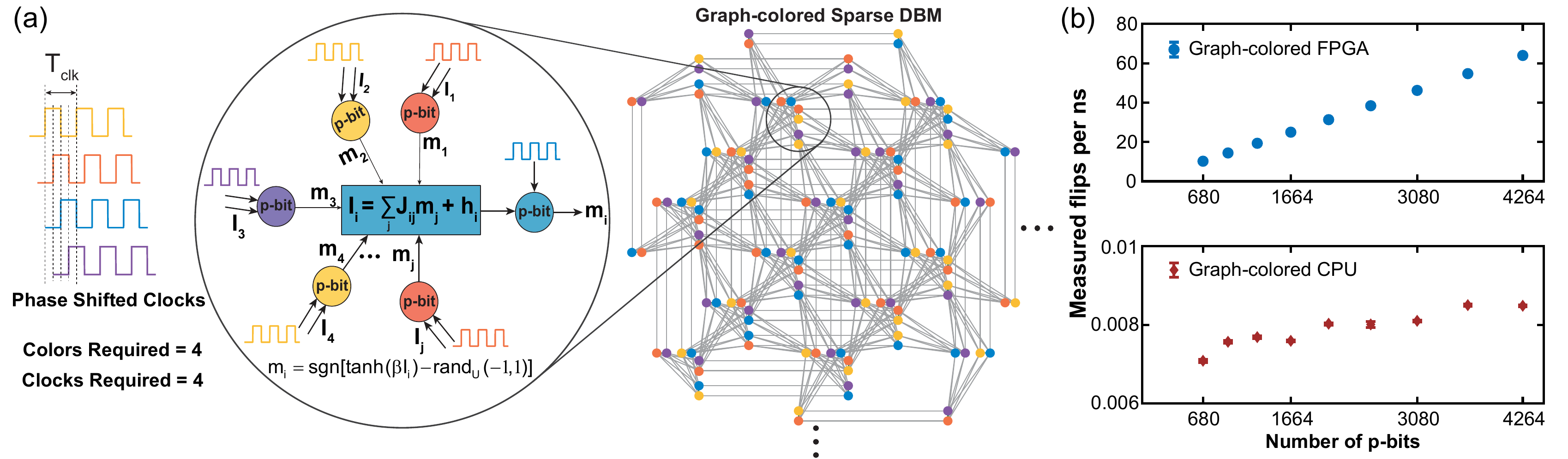}
\vspace{-10pt} 
    \caption{{\footnotesize (a) An example of massively parallel architecture with four parallel same-frequency and equally phase-shifted clocks to trigger the colored p-bit blocks. The sparse DBM (Pegasus 4,264 p-bits) is colored with four colors using the graph-coloring algorithm to exploit parallel updating of unconnected p-bits and the input for each p-bit is computed using  Eq.~\eqref{eq:synapse}. (b) Measured flips/ns as a function of graph size (number of p-bits) showing ideal parallelism scaling linearly with the system size in the case of the graph-colored FPGA (top). The graph-colored CPU flips/ns as a function of the graph size (bottom).}} 
    \label{fig:architecture}
    \vspace{-8pt}
\end{figure*}
To reduce the graph distance between the label bits, visible and hidden bits, we chose 5 sets of label bits (5$\times$10 = 50 p-bits) using one-hot encoding per digit. Experiments with more label bits did not show significant differences.  Also, experiments with multiple label bits in the RBM did not show any difference. This suggests that randomization of indices is particularly important for sparse models, but is unnecessary for fully-connected networks whose graph distance between any two nodes is a constant.

\section{p-computer Architecture}
\label{sec:architecture}

On the sparse DBM, we color the graph using the heuristic graph-coloring algorithm DSatur \cite{brelaz1979new} to exploit parallel updating of unconnected p-bits. This approach involves assigning different colors to connected p-bits and the same color to unconnected p-bits as shown in FIG.~\ref{fig:architecture}a to implement Gibbs sampling in a massively parallel manner on sparse and irregular graphs \cite{Aadit2022a}. Finding the minimum number of colors is an NP-hard problem, however, the minimum number of colors is not a strict requirement as sparse graphs require only a limited number of colors, and for our purpose, heuristic coloring algorithms like DSatur with polynomial complexity can color the graph efficiently.

In the case of the Pegasus graph with 4,264 p-bits, where the maximum number of neighbors is 15, only four colors are used as shown in FIG.~\ref{fig:architecture}a. Therefore we need four equally phase-shifted and same-frequency clocks for updating the p-bits in each color block one by one. Similarly, the Zephyr graph (3,360 p-bits and the maximum number of neighbors is 20) can also be colored with five colors using this procedure. In this approach, a graph comprised of $N$ p-bits is able to perform a full sweep in a single clock cycle ($T_{\text{clk}}$). We refer to this architecture as the pseudo-asynchronous Gibbs sampling \cite{chowdhury2023full}. The key advantage of this approach is that the p-computer becomes faster as the graph size grows as shown in FIG.~\ref{fig:architecture}b and Table~\ref{tab:fps} for both graph-colored FPGA and graph-colored CPU.

Parallelism offers many more samples to be taken at a clock cycle (scales as N, being the number of p-bits in the network as shown in FIG.~\ref{fig:architecture}b), however, we also establish that this parallelism does not introduce any approximations or errors by performing an ``inference'' experiment as discussed in Supplementary Section~\ref{sec:exactness_samples}.

\section*{Conclusion}
\label{sec:conclusions}
In this work, we have presented a hybrid probabilistic-classical computing setup to train sparse and deep Boltzmann networks. We used a sparse Ising machine with a massively parallel architecture that achieves a sampling speed orders of magnitude faster than traditional CPUs. Unlike similar hardware approaches where the MNIST dataset is downsampled, reduced, or replaced entirely for smaller datasets, we trained the full MNIST dataset without any simplifications. We also trained harder datasets like Fashion MNIST and a reduced version of CIFAR-10 with a grayscale scheme. Our sparse DBM matched the accuracy of RBMs while using 100$\times$ fewer parameters and successfully generated new images while the RBM failed to do so. We systematically studied the mixing time of the hardware-aware network topologies and showed that the classification accuracy of our model is not limited by the computational tractability of the algorithm but limited by the moderately sized FPGAs that we were able to use in this work. 

Further improvements may involve using deeper, wider, and perhaps ``harder to mix'' network architectures using more complex datasets, the use of higher-order energy functions beyond quadratic interactions \cite{sejnowski1986higher} and the use of Gaussian visible units to better model continuous data \cite{liao2022gaussian}. The use of higher-order interactions in the context of combinatorial optimization has generated significant attention \cite{bashar2023designing,bybee2023efficient,he2023many}, and its application to learning problems could be possible due to the simplified nature of multiplication with binary neurons. 
Moreover, combining layer-by-layer training techniques of conventional DBMs with our approach could lead to further possible improvements. Nanodevice implementations of such sparse Ising machines, for example, using stochastic magnetic tunnel junctions may significantly change this picture, potentially changing established wisdom on the practical use of deep Boltzmann networks. 

\appendix

\section*{Methods}

\subsection{FPGA and CPU specifications}
\label{sec:fpga_cpu_spec}
In this article, Xilinx Alveo U250, a data center accelerator card (Virtex UltraScale+ XCU250 FPGA) with peripheral component interconnect express (PCIe) connectivity has been used \cite{xilinx-u250}. PCIe interface performs data transfer at the rate of 2.5 gigatransfers per second (GT/s). The classical computer used in this study is equipped with an 11th Gen Intel Core i7-11700 processor with a clock speed of up to 4.90 GHz and 64 GB of random access memory (RAM).

The digital implementation of p-bits consists of a pseudorandom number generator (PRNG), a lookup table for the activation function (tanh), and a threshold to generate a binary output (details in the Supplementary Section~\ref{sec:fpga}). The read-out architecture with mirror p-bits is discussed in the Supplementary Section~\ref{sec:read_out}. Weights and biases with fixed point precision of 10 bits (1 sign bit, 6 integer bits, and 3 fraction bits) are used to provide tunability through the activation function. 

\subsection{MNIST data, D-Wave graphs and RBM code}
\label{sec:mnist_data}
MNIST files are downloaded from \cite{lecun1998mnist}. Then the image data are converted to binary form (black and white) by rounding up the pixel intensities in MATLAB. While we focused on black and white images for our main results, in the Supplementary Section~\ref{sec:fashionMnist}-\ref{sec:cifar100}, we show how the learning algorithm can be extended to learn grayscale images, following a similar time-averaging approach discussed in Ref.~\cite{hirtzlin2019stochastic}. The Pegasus and Zephyr graphs are extracted using the procedure described in \cite{dwave-docs}. The RBM code used in this work is similar to the one available in \cite{hinton2014training}.

\subsection{Data transfer between FPGA and CPU}
\label{sec:data_transfer}
A PCIe interface is used to communicate between FPGA and CPU through MATLAB interface for the `read/write' operations (see Supplementary FIG.~\ref{fig:fpga_imp}c). A global `disable/enable' signal broadcast from MATLAB to the FPGA is used to freeze/resume all p-bits. Before a `read' instruction, the p-bit states are saved to the local block memory (BRAM) with a snapshot signal. Then the data are read once from the BRAM using the PCIe interface and sent to MATLAB for post-processing i.e., computing gradients and updating the weights. For the `write' instruction, the `disable' signal is sent from MATLAB to freeze the p-bits before sending the updated weights. After the `write' instruction is done, p-bits are enabled again with the `enable' signal sent from MATLAB. The data transfer efficiency is influenced by this back-and-forth communication between the FPGA and MATLAB. Furthermore, the conversion of bipolar to binary weights and biases during each epoch (as explained in Supplementary~\ref{sec:bipolar_to_binary}) adds some time overheads while sending them from MATLAB to FPGA. Even though sampling is very fast in FPGA, due to these overheads it takes $\approx$\,20 hours to train full MNIST on 4,264 p-bit Pegasus with CD-$10^5$ for 100 epochs. This issue can be improved significantly by updating the weights and biases inside the FPGA. To understand the improvement introduced by our hybrid approach, we also note that the corresponding equivalent version with CD-$10^4$ and with the same graph coloring on a CPU took $\approx 2.4$ days (57.5 hours) to complete only 10 epochs (projected time for 100 epochs is $\approx 24$ days).    

\subsection{Measurement of flips per nanosecond}
\label{sec:fps_measure}
To measure the flips/ns, one p-bit in each color block is designed with a programmable counter in the FPGA to count the flip attempts. A reference counter running parallelly is set to count up to a preset value at the positive edge of a reference clock. When the reference counter is done counting, the p-bit counters are stopped. Comparing the p-bit counter outputs (representing the total number of attempted flips in each color block) with the reference counter preset value, the time for the total flips is obtained. With this data, the flips/ns of the p-computer is measured experimentally. To determine the flips/ns for the standard CPU and graph-colored CPU, MATLAB built-in `tic' and `toc' functions are used to measure the elapsed time while counting the total flips. The flips/ns is measured in real-time using this data. The error bars in FIG.~\ref{fig:architecture}b are obtained by taking 500 measurements of flips/ns.

\section*{Acknowledgements}
We gratefully acknowledge discussions with Dr. Jan Kaiser. We are thankful to the Xilinx University Donation Program (XUP) for the FPGA development boards and G. Eschemann for useful discussions on airhdl. This work is partially supported by an Office of Naval Research Young Investigator Program grant and a National Science Foundation CCF 2106260 grant.

\section*{Data availability}
The data that support the plots within this paper and other findings of this study are available from the corresponding author upon reasonable request.

\section*{Code availability}
The computer code used in this study is available from the corresponding author upon reasonable request.

\section*{Author contributions}
SN and KYC conceived the study. KYC supervised the study. SN and NAA developed the hybrid FPGA-CPU implementation. SN and SC performed the benchmark RBM training. SN and NAA performed the FPGA experiments to train sparse DBMs. SN, NAA, MM, SC, YQ, KYC discussed, analyzed the experiments, and participated in writing the manuscript. 
\section*{Competing interests}
The authors declare no competing interests.

\balance{

\bibliographystyle{unsrtnat}}

\clearpage

\onecolumngrid
\section*{Supplementary Information}
\beginsupplement

\setcounter{subsection}{0}
\subsection{FPGA Implementation}
\label{sec:fpga}

We present the experimental results in the main paper using a hybrid classical-probabilistic computer. Here, we discuss the technical details of the FPGA-based p-computer implemented on a Xilinx Alveo U250 data center accelerator card. The basic architecture is presented in FIG.~\ref{fig:fpga_imp}. 

\subsubsection{p-bit and MAC Unit}
\label{pbit_MAC}
\begin{figure*}[h]
    \centering
    \includegraphics[width =1 \textwidth ]{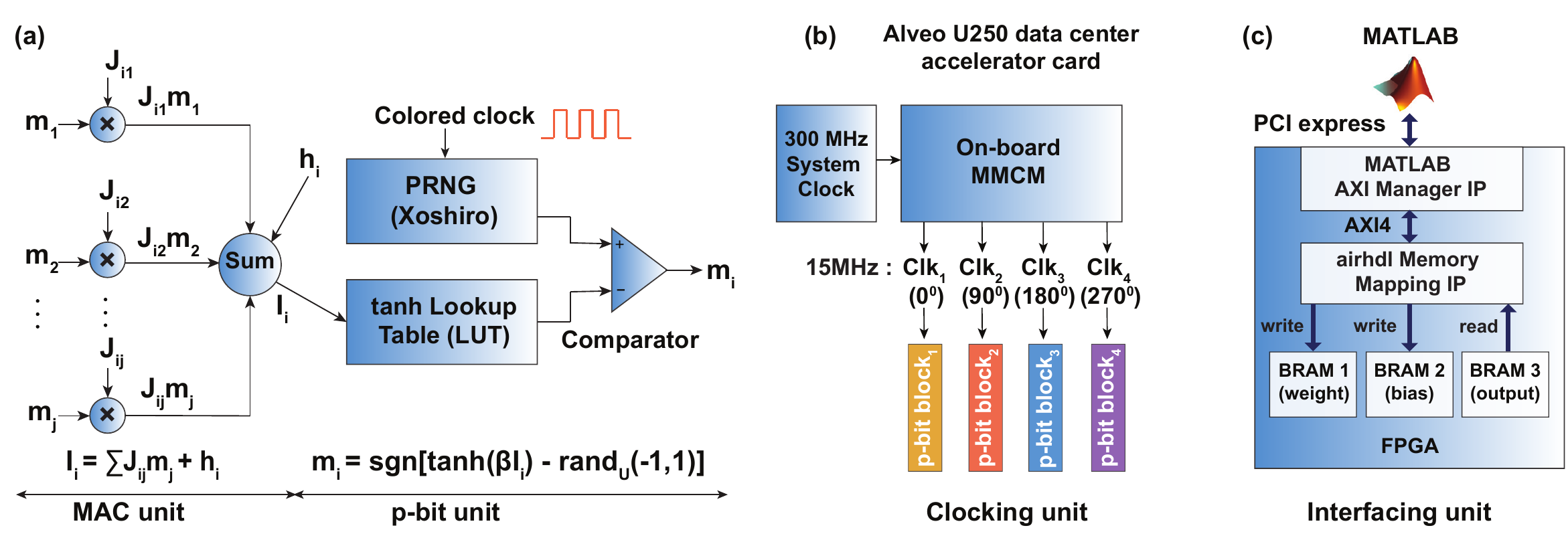}
    \caption{(a) The MAC (multiplier–accumulator) unit implements Eq.~\eqref{eq:synapse}. The p-bit unit consists of a xoshiro pseudorandom number generator (PRNG), a lookup table for the activation function (tanh), and a comparator to generate a binary output. (b) A built-in clocking unit generates equally phase-shifted and same-frequency parallel clocks to trigger the PRNGs inside the colored p-bit blocks. (c) A PCIe interfacing unit transfers data between MATLAB and the FPGA.}
\label{fig:fpga_imp}
\vspace{-15pt}
\end{figure*}

 A single p-bit consists of a tanh lookup table (LUT) for the activation function, a pseudorandom number generator (PRNG), and a comparator to implement  Eq.~\eqref{eq:pbit}. Digital input with a fixed point precision (10 bits with 1 sign, 6 integers, and 3 fraction bits) provides tunability through the activation function. The effect of different fixed point precisions has been explored in Supplementary Section~\ref{sec:bit_precision}. In this work, we used a high-quality 32-bit PRNG (xoshiro \cite{blackman2021scrambled}). There is also a multiplier–accumulator (MAC) unit to compute  Eq.~\eqref{eq:synapse} based on the neighbor p-bit states and provide input signal for the LUT as shown in FIG.~\ref{fig:fpga_imp}a. 

 \subsubsection{Bipolar to binary conversion}
\label{sec:bipolar_to_binary}
As described in Section~\ref{sec:hybrid} of the main article, Eq.~\eqref{eq:en}-\eqref{eq:pbit} are represented with the bipolar state of the variables. However, for the FPGA-based implementation of p-computer, using a binary state of variables is more convenient since digital CMOS operates with gnd (0) and VDD (1) and Boolean logic can naturally represent the p-bit state. Therefore, bipolar to binary conversion is done on the weights and biases before being sent to the FPGA, according to the following mapping:  
\begin{eqnarray}
{J}_{\text {binary }}&=&2 {J}_{\text {bipolar }}\\
{h}_{\text {binary }}&=&{h}_{\text {bipolar }}-{J}_{\text {bipolar }} \mathbf{1}
\end{eqnarray}
where, $\mathbf{1}$ is a vector of ones of size $N\times1$, and $N$ is the number of p-bits. All the variables in the MAC unit are also calculated in binary notations. To convert the stored activation function values of LUT from bipolar to binary representation, the $\mathrm {tanh(x)}$ is mapped to $ {[1 + \tanh{(x)}]/2}$. Finally, the output p-bit is represented with binary states $m_i \in \{0, 1\}$.

\subsubsection{Clocking unit}
\label{sec:clock_unit}
In Section~\ref{sec:architecture} of the main paper, we discussed graph coloring to color p-bit blocks achieving massive parallelism, which we define as the linear scaling of probabilistic flips/ns with respect to increasing graph size. To achieve this parallelism, we utilized multiple built-in clocks inside the FPGA board to drive the PRNGs within the p-bit blocks, as shown in FIG.~\ref{fig:fpga_imp}a. The mixed-mode clock manager (MMCM) block, available in the Xilinx Alveo U250 data center accelerator card, generates equally phase-shifted and same-frequency parallel stable clocks. The input to the MMCM is a 300 MHz differential pair system clock created on a low-voltage differential signaling (LVDS) clock-capable pin (FIG.~\ref{fig:fpga_imp}b). These clocks are highly precise with minimal noise or phase deviation. By triggering the colored p-bit blocks with these phase-shifted clocks, we achieve massive parallelism.

\subsubsection{Interfacing unit}
\label{sec:interface}
We use MATLAB as an Advanced eXtensible Interface (AXI) manager to communicate with the FPGA board through a PCI express interface (FIG.~\ref{fig:fpga_imp}c). We have designed an AXI manager integrated IP on the board that transfers data with a 32-bit memory-mapped slave register IP via the fourth-generation AXI (AXI4) protocol. An external website `airhdl' \cite{airHDL} is used to manage the memory mapping of the registers into the block rams (BRAMs) inside the FPGA. We used two BRAMs to write the weights and biases from MATLAB and another BRAM to save the p-bit states that MATLAB reads out.

\subsection{Readout architecture with mirror p-bit}
\label{sec:read_out}

\begin{figure*}[h]
    \centering
    \includegraphics[width=1.0\textwidth]{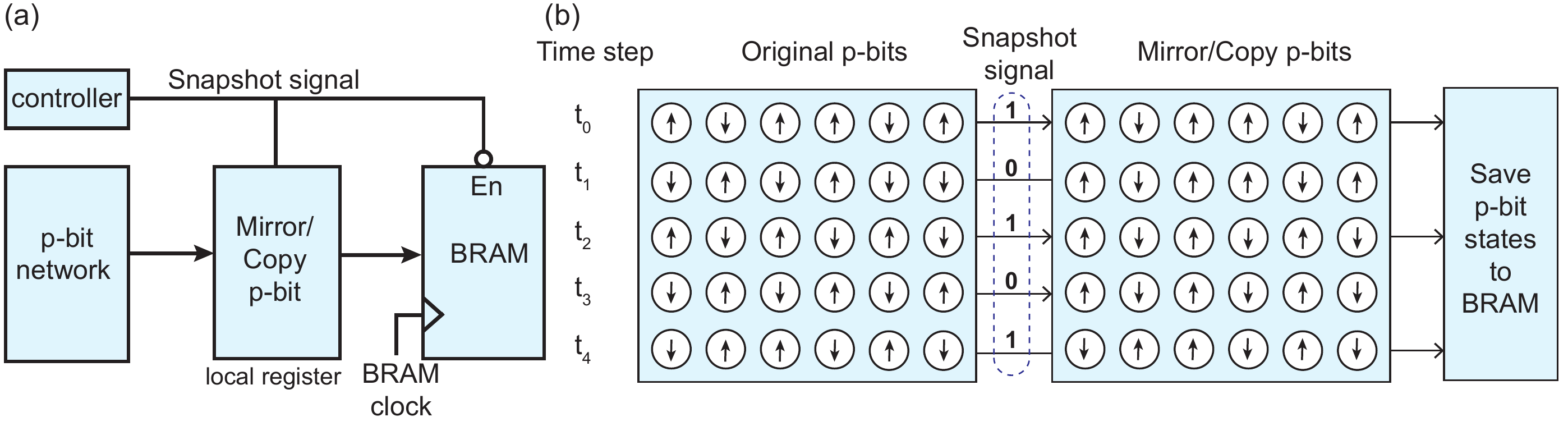}
    \caption{{\footnotesize (a) Block diagram showing the mirror (or copy) p-bit architecture with a snapshot signal. The controller block generates the snapshot signal at which time the original p-bit states are copied to local memory (registers) and at the inverted snapshot signal those states are saved into block memory (BRAM), only once. Subsequent zero signals from the snapshot signal do nothing to mirror/copy p-bits or to BRAM. (b) Conceptual diagram to visualize the operation of the snapshot signal.}}
    \label{fig:mirror_pbit}
\end{figure*}

FIG.~\ref{fig:mirror_pbit} shows our readout architecture that can take a ``snapshot'' of the system at a given time. The nature of the learning algorithm is that only a handful of equilibrium samples are needed to estimate correlations and averages without requiring continuous samples. By way of mirror (or copy) p-bit registers, we decouple system p-bits (that are always on) from snapshot states that are saved to memory. In this way, while the system goes through a large number of samples, we are able to take a sufficient number of samples that are saved to local memory (at our own accord) which can then be read by a CPU.

\begin{figure*}[b!]
    \centering
    \includegraphics[width=0.65\textwidth]{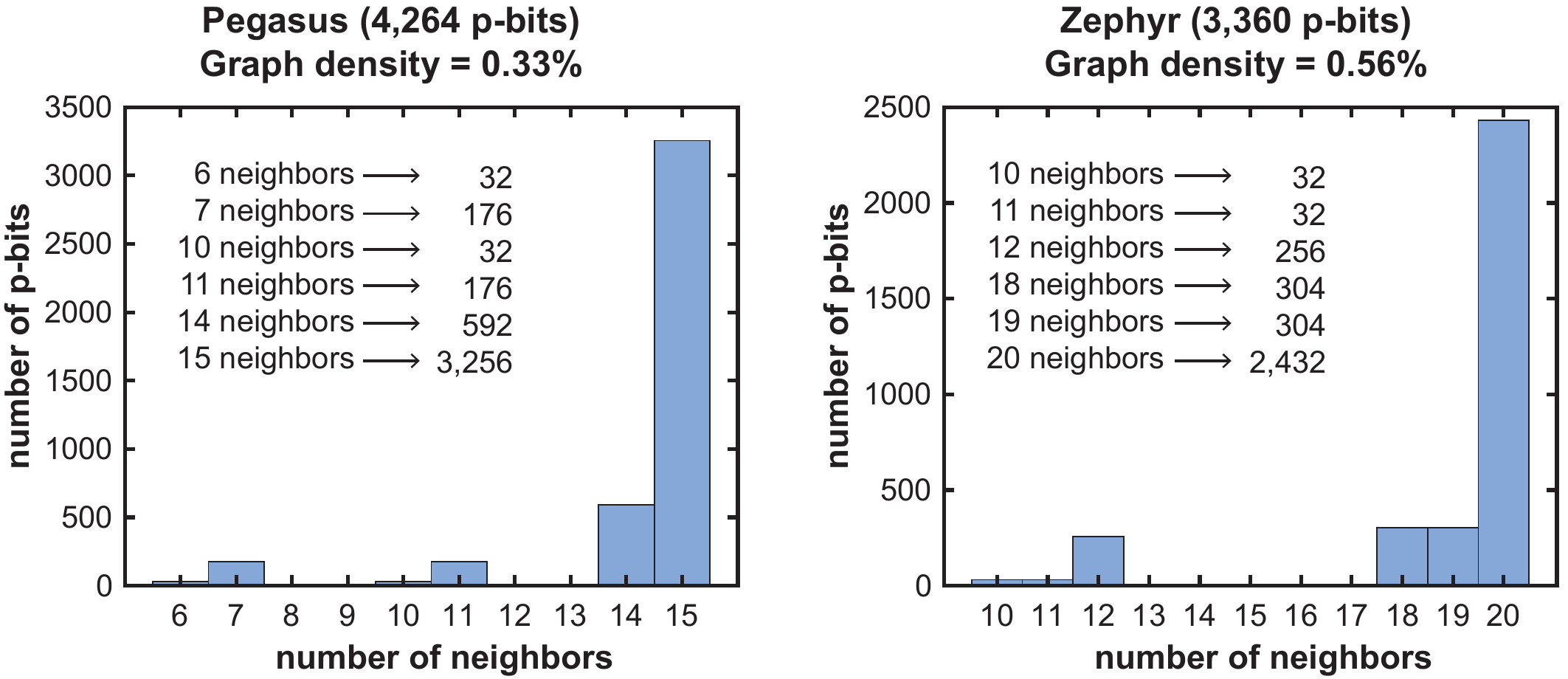}
    \caption{{\footnotesize The graph density and the neighbor distribution of sparse DBMs (Pegasus and Zephyr) where graph density, $\rho = \displaystyle 2|E| /(|V|^2-|V|)$ where $\displaystyle|E|$ = the number of edges and $\displaystyle|V|$  = the number of vertices in the graph. The density of Pegasus 4,264 p-bits is 0.33\% and 3,256 p-bits have the maximum number of neighbors 15. Zephyr (3,360 p-bits) has a density of 0.56\% and 2,432 p-bits have the 20 maximum neighbors.}}
    \label{fig:sparsity_dbms}
\end{figure*} 

\subsection{Graph density of sparse DBMs}
\label{sec:sparsity_dbm}
The networks we present in the manuscript (Pegasus 4,264 p-bits and Zephyr 3,360 p-bits) are highly sparse, which we quantitatively show in FIG.~\ref{fig:sparsity_dbms}.
We use the typical graph density metric, measured by
\begin{equation}
\mbox{graph density} = \frac{\mbox{number of edges in the network}}{\mbox{number of edges in a fully-connected network}}\times 100 
\end{equation}
By this metric, the network we have shown in FIG.~\ref{fig: overview}e (Pegasus, 4264) only has a graph density of $0.33\%$. Moreover, we also show a vertex degree distribution of the network providing a histogram of nodes and the number of their neighbors. On the right, the same metrics are shown for the Zepyhr graphs with similar results.

\subsection{Training accuracy of MNIST/100}
\label{sec:train_100im}
\begin{figure*}[h]
    \centering
    \includegraphics[width =.35 \textwidth ]{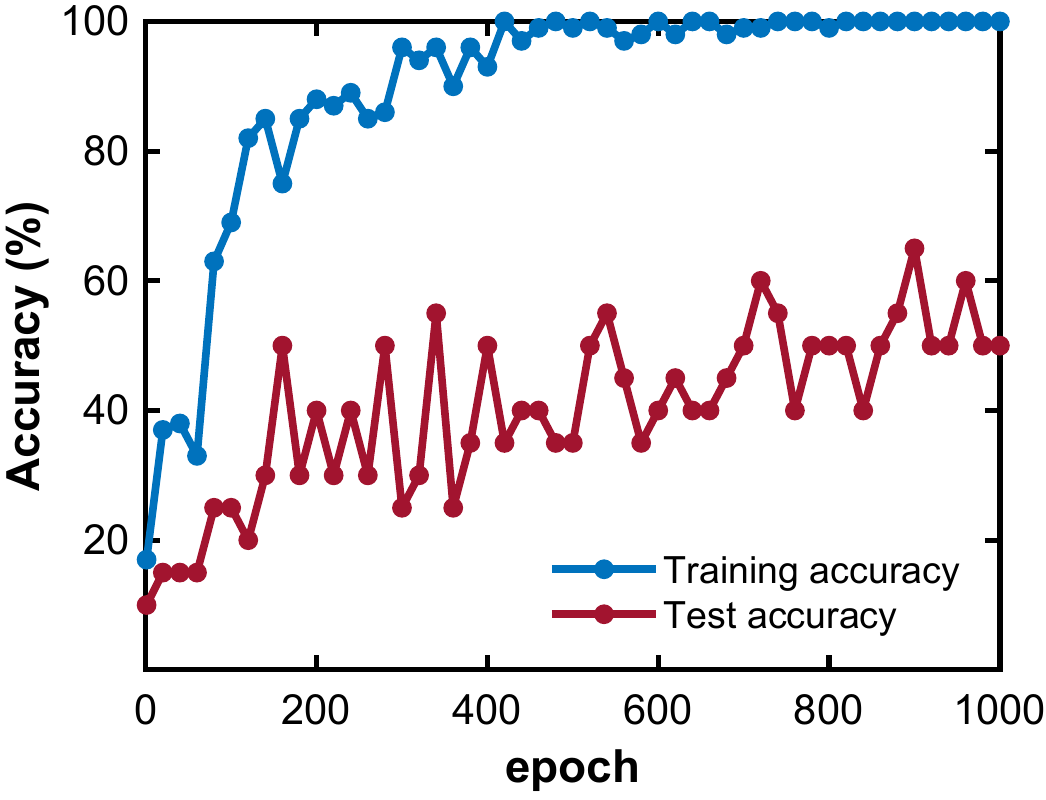}
    \caption {Accuracy of training 100 images with sparse DBMs up to 1,000 epochs. Training is accomplished with 10 mini-batches and CD-{$10^5$}. All the trained 100 images and 20 unseen images have been tested here.}
\label{fig:acc_100im}
\end{figure*}

We have studied the effect of training the sparse DBMs (Pegasus 4,264 p-bits) with a small subset of data before training the full MNIST. In this setting, we chose 100 images from MNIST to train on the sparse network with our massively parallel architecture. To train these 100 images, we used 10 mini-batches, having 10 images in each batch and the same set of hyperparameters as in training full MNIST. The training accuracy reached 100\% within 1,000 epochs as illustrated in FIG.~\ref{fig:acc_100im}. We also explored different values of the `regularization' parameter ($\lambda$ = 0.005, 0.001) which is generally used to keep the weights from growing too much. In our case of sparse DBM, we did not observe any significant difference between a small regularization and without any regularization. 
The poor test accuracy here is an indication of overfitting due to the small size of the training set (only 100 images). We observed similar accuracy on Zephyr graphs with 3,360 p-bits for 100 images. 

\subsection{Effect of weight precision}
\label{sec:bit_precision}
\begin{figure*}[h]
    \centering
    \includegraphics[width =.7\columnwidth ]{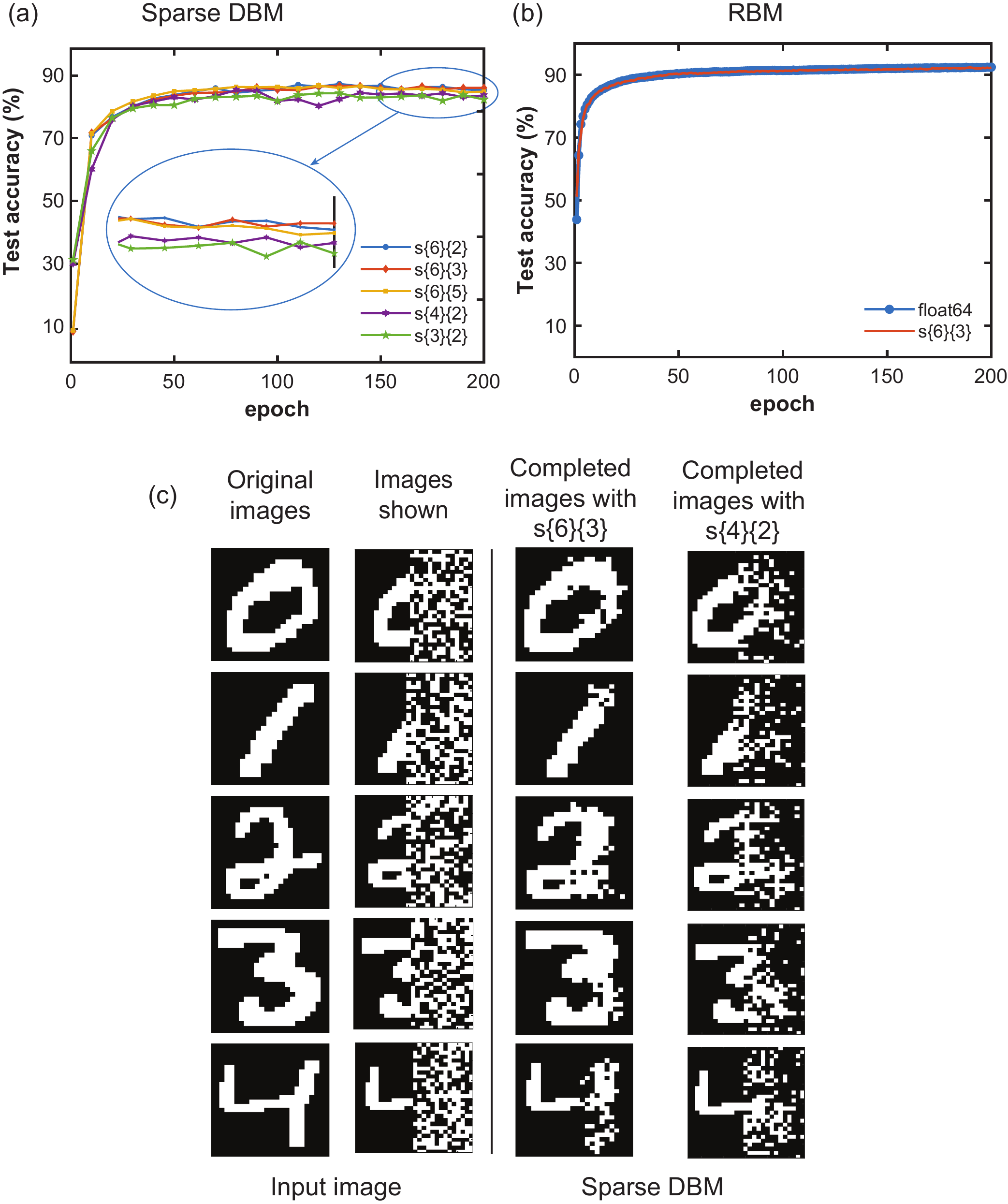}
    \caption {(a) Test accuracy of full MNIST with sparse DBMs (Pegasus 2,560 p-bits) up to 200 epochs with five different fixed point precisions of weights (s\{6\}\{2\}, s\{6\}\{3\}, s\{6\}\{5\}, s\{4\}\{2\} and s\{3\}\{2\}). (b) Test accuracy of full MNIST for RBM (512 hidden units) with double-precision floating point 64 bits and s\{6\}\{3\}. (c) Image completion examples with sparse DBMs for fixed point precisions of weights s\{6\}\{3\} and s\{4\}\{2\} (annealing schedule varies from $\beta$\,=\,0 to  $\beta$\,=\,5 with 0.125 steps). With s\{6\}\{3\} precision, the network can complete the images where the right half of the image starts from random noise. Below s\{4\}\{2\}, the network fails to complete the images.}
\label{fig:bit_precsn}
   
\end{figure*}

The results we have in Section~\ref{sec:train} in the main paper utilized a weight precision of 10 bits (1 sign, 6 integer, and 3 fraction bits i.e., s\{6\}\{3\}). Here, we explore different weight precisions by changing the fraction bit width and compare the results to identify the effect of weight precision on accuracy. We trained full MNIST with 1,200 mini-batches for 200 epochs, using five different weight precisions of s\{6\}\{2\}, s\{6\}\{3\}, s\{6\}\{5\}, s\{4\}\{2\} and s\{3\}\{2\}. We chose a Pegasus 2,560 p-bit graph as a sparse DBM for this experiment that fits into the FPGA since increasing bit precision reduces the available resources. The weight update is accomplished in MATLAB (double-precision floating point with 64 bits), but before the new weights are loaded to the FPGA, they are converted to the corresponding fixed-point precision. The choice of hyperparameters remains the same for all cases. The test accuracy goes to $\approx88$\% in each case (with 17,984 parameters) and there is no remarkable difference among the accuracy of the different weight precisions between s\{6\}\{5\} to s\{6\}\{3\}, accuracy starts degrading at or below s\{4\}\{2\} (FIG.~\ref{fig:bit_precsn}a).  We also trained full MNIST on RBM (512 hidden units) using both float64 and s\{6\}\{3\} weight precision for 200 epochs. The test accuracy remains the same for these two different precisions as shown in FIG.~\ref{fig:bit_precsn}b.

To further study the impact of weight precision on the \emph{generative} properties of the sparse DBM network, we have conducted image completion experiments. FIG.~\ref{fig:bit_precsn}c shows inference experiments where we obscure half of an image and let a trained network evolve to the corresponding minima (by annealing the network from $\beta$\,=\,0 to  $\beta$\,=\,5). We observe that while s\{6\}\{3\} can complete this task, precisions below s\{4\}\{2\} start failing.

\subsection{Mixing times in Pegasus graph (3,080 p-bits)}
\label{sec:mixing_P3080}
We described the mixing times in Section~\ref{sec:equilibrium} of the main paper showing the results (FIG.~\ref{fig:real_sweep}) from our largest size Pegasus (4,264 p-bits).  Here, we show another graph, Pegasus with 3,080 p-bits to measure the mixing time of the network. Unlike the main model, for this experiment, we trained full MNIST for only 50 epochs (instead of 100) using the same hyperparameters as mentioned in the main Section~\ref{sec:train} with different numbers of sweeps starting from CD-$10^{2}$ to CD-$10^{6}$. Test accuracy improves significantly when we take more than CD-$10^{4}$ per epoch.

\begin{figure}[h]
    \centering
    \includegraphics[width =.75 \columnwidth ]{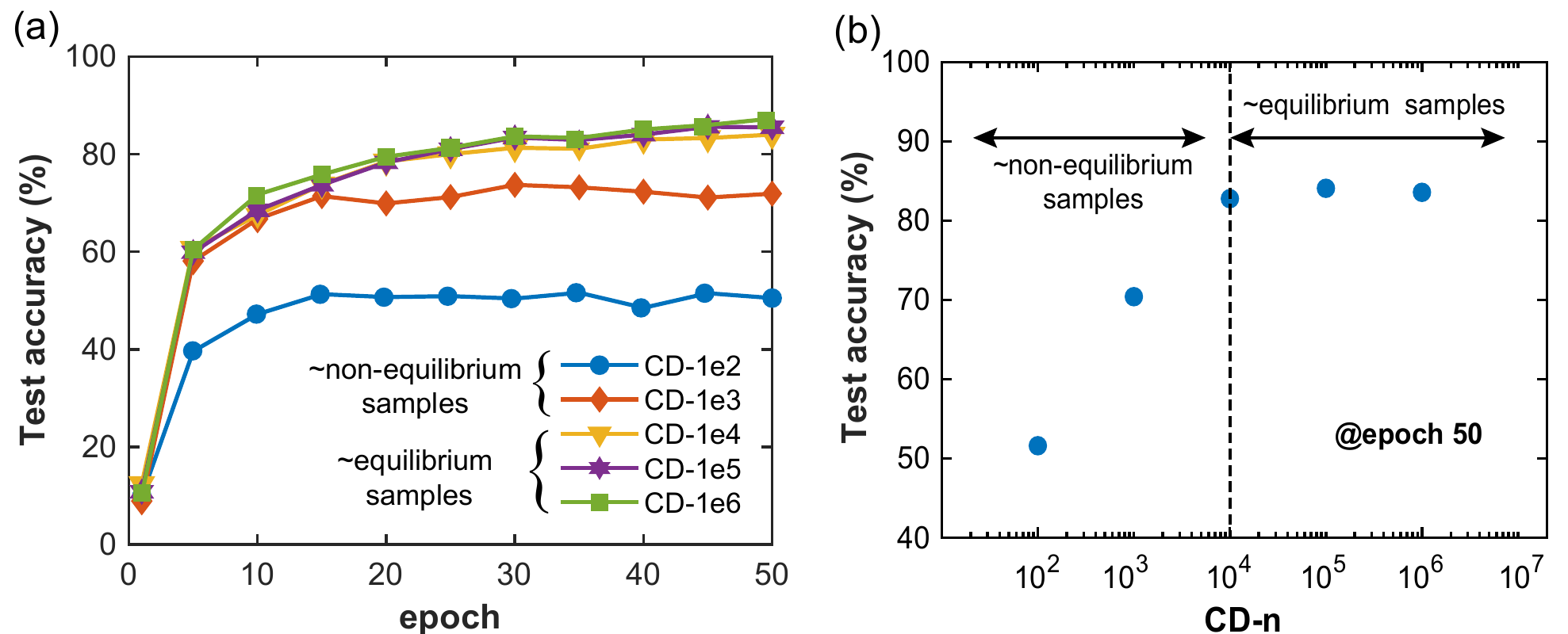}
    \caption {(a) Test accuracy after training full MNIST up to 50 epochs with different numbers of sweeps using sparse DBMs (Pegasus 3,080 p-bits). For our sparse graph, to mix the Markov chain properly we need minimum CD-$10^{4}$. Reducing the number of sweeps significantly degrades the quality of mixing in the chain. (b) Test accuracy as a function of CD-n at epoch 50 showing the equilibrium and non-equilibrium samples.}
\label{fig:sweep_P3080}
   
\end{figure}

\subsection{Image generation with Zephyr}
\label{sec:ImSynth_zephyr}
\begin{figure*}[h]
    \centering
    \includegraphics[width =.63 \textwidth ]{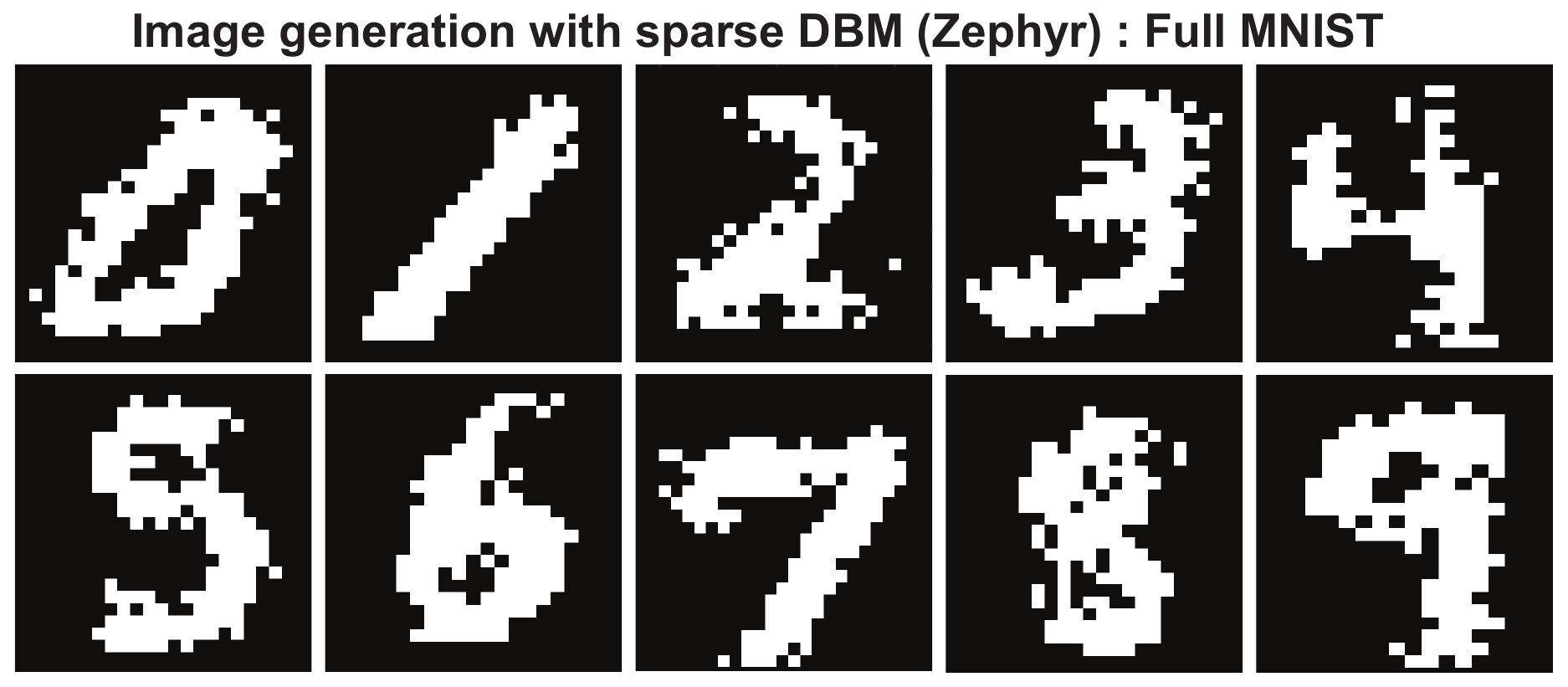}
    \caption {Image generation examples with sparse DBM Zephyr (3,360 p-bits) after training the network with the full MNIST dataset.}
\label{fig:zephyr}
\end{figure*}

In the main paper, FIG.~\ref{fig: overview}f displays the images generated with Pegasus (4,264 p-bits) graph and the procedure is described in Section~\ref{sec:ImageSynth}. Here we explored image generation with a different type of sparse DBM, Zephyr (3,360 p-bits) that also reached $\approx90$\% accuracy with randomized indices as demonstrated in the main FIG.~\ref{fig:rand_label}c (bottom). The generated images with Zephyr as shown in FIG.~\ref{fig:zephyr}  are slightly different from the Pegasus ones.

\subsection{Momentum to the learning rule}
\label{sec:momentum}

In our training, we used the momentum in our update rules, which are empirically added to the learning rule we discuss in the next section. By retaining a portion of the last update to the weights, momentum helps increase the effective learning rate \cite{hinton2012practical}. The effective increase in the learning rate is equivalent to multiplying it by a factor of 1/(1-$\alpha$) where $\alpha$ is denoted as momentum. Using this process, the algorithm can increase the effective learning rate without causing unstable oscillations, which ultimately speeds up the convergence of the training process \cite{tieleman2008training}. We modify the learning rule equations in the main Eq.~(\ref{eq:del_J}) and Eq.~(\ref{eq:del_h})  by introducing the momentum term as follows:
\begin{eqnarray}
   \Delta J_{ij}(n)& =& \varepsilon\bigg(\langle m_im_j\rangle_\text{{data}}-\langle m_im_j\rangle_\text{{model}}\bigg) + \alpha \Delta J_{ij}(n-1) \label{eq:momentum_J}\\
   \Delta h_{i}(n) &=& \varepsilon\bigg(\langle m_i\rangle_\text{{data}}-\langle m_i\rangle_\text{{model}}\bigg) + \alpha \Delta h_{i}(n-1)
   \label{eq:momentum_h}
\end{eqnarray}
where $n$ represents the $j$\textsuperscript{th} index (ranging from 1 to the number of batches) in Algorithm 1 in the main paper. 

\subsection{Maximum Likelihood for Boltzmann Networks}
\label{sec:derivation}
The basic idea of Boltzmann networks is to start from a physics-inspired variational guess, that the data distribution will be approximated by a model whose probability $p^M$ for a given input vector $m^{(i)}$ ($i$, being the input index) obeys the Boltzmann law (ignoring biases in our derivation for simplicity): 
\begin{equation}
    p^M \left(m^{(i)}\right) =\frac{1}{Z} \exp \left[\sum_{e r} J_{e r} m_e^{(i)} m_r^{(i)}\right]
    \label{eq:bm}
\end{equation}

In our setting, we have a system of $M$ fully visible p-bits connected in some arbitrary graph topology. The problem is learning a  ``truth table'' with exactly $N_T$ lines of inputs in it. The model is going to try to select these $N_T$ states in the space of $2^M$ possible discrete probabilities. Like in any other ML model, fitting every line of the truth table exactly will overfit, but the Boltzmann formulation given by Eq.~(\ref{eq:bm})  smooths out the sum of ``delta function''-like data vectors in the $2^M$ space, which can later be used for generating new samples.  

We define a $p^V$ as the probability distribution of the data, corresponding to the visible bits. Then, a maximum likelihood estimation minimizing the Kullback–Leibler divergence between the data and the model can be used to derive the learning rule, by taking the negative derivative of $KL(p^V \| \ p^M)$:

\begin{equation}
    KL(p^V \| \ p^M)=\sum_{i=1}^{N_T } p_{i}^{V}\log\bigg[\frac{p_{i}^{V}}{p_{i}^{M}}\bigg] \label{kl_eqn}
\end{equation}
where $i$ is the index of truth table lines $N_T$. To simplify analysis, we consider fully visible networks where $p_{i}^{V}$ is independent of $J_{mn}$ for any network topology since $p_{i}^{V}$ represents the data distribution. 

\begin{equation}
\frac{\partial K L\left(p^V \| \ p^M\right)}{\partial J_{m n}}=\frac{\partial p_i^V}{\partial J_{m n}} \log \left[\frac{p_i^V}{p_i^M}\right]=\sum_{i=1}^{N T} \frac{p_i^V}{p_i^M}\left(\frac{-\partial p_i^M}{\partial J_{m n}}\right) \label{kl_partial}
\end{equation}
\begin{equation}
\frac{\partial p_i^M}{\partial J_{m n}}=\frac{\partial}{\partial J_{m n}}\left[\frac{1}{Z} \exp \left[\sum_{e r} J_{e r} m_e^{(i)} m_r^{(i)}\right]\right] \label{piM_partial}
\end{equation}

 \begin{equation}
Z=\sum_{j=1}^{2^N} \exp \left[\sum_{e r} J_{e r} m_e^{(j)} m_r^{(j)}\right] \label{Z_eqn}
\end{equation}

where the index $j$ represents all possible states from 1 to $2^M$ for the model.
\begin{equation}
\begin{aligned}
\frac{\partial p_i^M}{\partial J_{m n}} & =\frac{-1}{Z^2} \cdot \frac{\partial Z}{\partial J_{m n}} \cdot\left[\exp \sum_{e r} J_{e r} m_e^{(i)} m_r^{(i)}\right]+\frac{1}{Z}\cdot \frac{\partial}{\partial J_{m n}} \left[\exp \sum_{e r} J_{e r} m_e^{(i)} m_r^{(i)}\right] \\
& =\frac{-\partial Z}{\partial J_{m n}} \cdot \frac{1}{Z} \cdot p_i^{M}+m_m^{(i)} \cdot m_n^{(i)} \cdot p_i^M \\
& =-\sum_{j=0}^{2^N-1} m_m^{(j)} m_n^{(j)} \cdot p_j^M \cdot p_i^M+m_m^{(i)} m_n^{(i)} \cdot p_i^M
\end{aligned} \label{final_piM}
\end{equation}
\begin{equation}
\begin{aligned}
-\frac{\partial K L\left(p^V \| \ p^M\right)}{\partial J_{m n}}&=\sum_{i=1}^{N_T} \frac{p_i^V}{p_i^M} \cdot \frac{\partial p_i^M}{\partial J_{m n}} \\
& =\sum_{i=1}^{N_T} \frac{p_i^V}{p_i{ }^M}\left[m_m^{(i)} \cdot m_n^{(i)} \cdot p_i^M-\sum_{j=0}^{2^M-1} m_m^{(j)} \cdot m_n^{(j)} \cdot p_j^M \cdot p_i^M\right] \\
& =\left[\sum_{i=1}^{N_T} p_i^V \cdot m_m^{(i)} m_n^{(i)}-\sum_{i=1}^{N_T} p_i^V \sum_{j=0}^{2^M-1} m_m^{(j)} \cdot m_n^{(j)} \cdot p_j^M\right] \\
& =\left[\left\langle m_m m_n\right\rangle_{\text {data }}-\left\langle m_m m_n\right\rangle_{\text {model }}\right] \\
&
\end{aligned}
\end{equation} 
which gives the familiar learning rule. A similar learning rule in terms of the averages can be derived by accounting for the biases in the energy, which we ignored for simplicity.

\subsection{Comparison with state-of-the-art GPUs and TPUs }
\label{sec:gpu} 
In the main paper, we show how graph-colored FPGA achieves massive parallelism to provide a few orders of magnitude faster sampling throughput than traditional CPUs. Here in Supplementary Table~\ref{tab:benchmarking}, we also compare the sampling speed to some state-of-the-art (SOTA) Ising machines implemented on the latest GPUs and TPUs. The throughput reported in this work up to 64 billion flips per second outperforms the numbers reported by the SOTA Ising solvers in GPUs and TPUs. It is also important that this comparison is not completely accurate and favors the GPU/TPU implementations for two reasons: First, all the GPUs and TPUs discussed here are solving simple, nearest-neighbor chessboard lattices, unlike the irregular and relatively high-degree (with up to 20 neighbors) graphs used in this work. Second, GPU/TPU implementations generally use low precision \{+1,-1\} weights (compared to 10 bits of weight precision in our work) and thus can explore only a few discrete energy levels. Both of these features are heavily exploited in reporting a large degree of flips/ns in these solvers and their performance would presumably be much worse if they were implemented in the same graphs with the same precision we discuss in this work.

\begin{table*}[h]
    \centering
     \caption{{\footnotesize Optimized GPU and TPU implementations of Markov Chain Monte Carlo sampling with regular chessboard lattices. It is important to note that these TPU and GPU implementations solve Ising problems in sparse graphs, however, their graph degrees are usually restricted to 4 or 6, unlike more irregular and higher degree graphs.}}
     \vspace{4pt}
    \begin{tabular}{@{}lcccc@{}}
        \toprule
        {\bf Sampling method} & {\bf topology} & {\bf max. degree} & {\bf flips/ns} \\ \midrule
        Nvidia Tesla C1060 GPU \cite{block2010multi, preis2009gpu}  & Chessboard  & 4 & 7.98 \\
        Nvidia Tesla V100 GPU \cite{yang2019high}  & Chessboard  & 4 & 11.37 \\
        Google TPU \cite{yang2019high} & Chessboard  & 4 & 12.88\\
        Nvidia Fermi GPU \cite{fang2014parallel} & Chessboard   & 4 &  29.85\\
        \bottomrule
         \end{tabular}
   \label{tab:benchmarking}
\end{table*}

\subsection{Image completion with DBM and RBM}
\label{sec:im_compltn_dbm_rbm}

\begin{figure*}[h]
    \centering
    \includegraphics[width=0.47\textwidth]{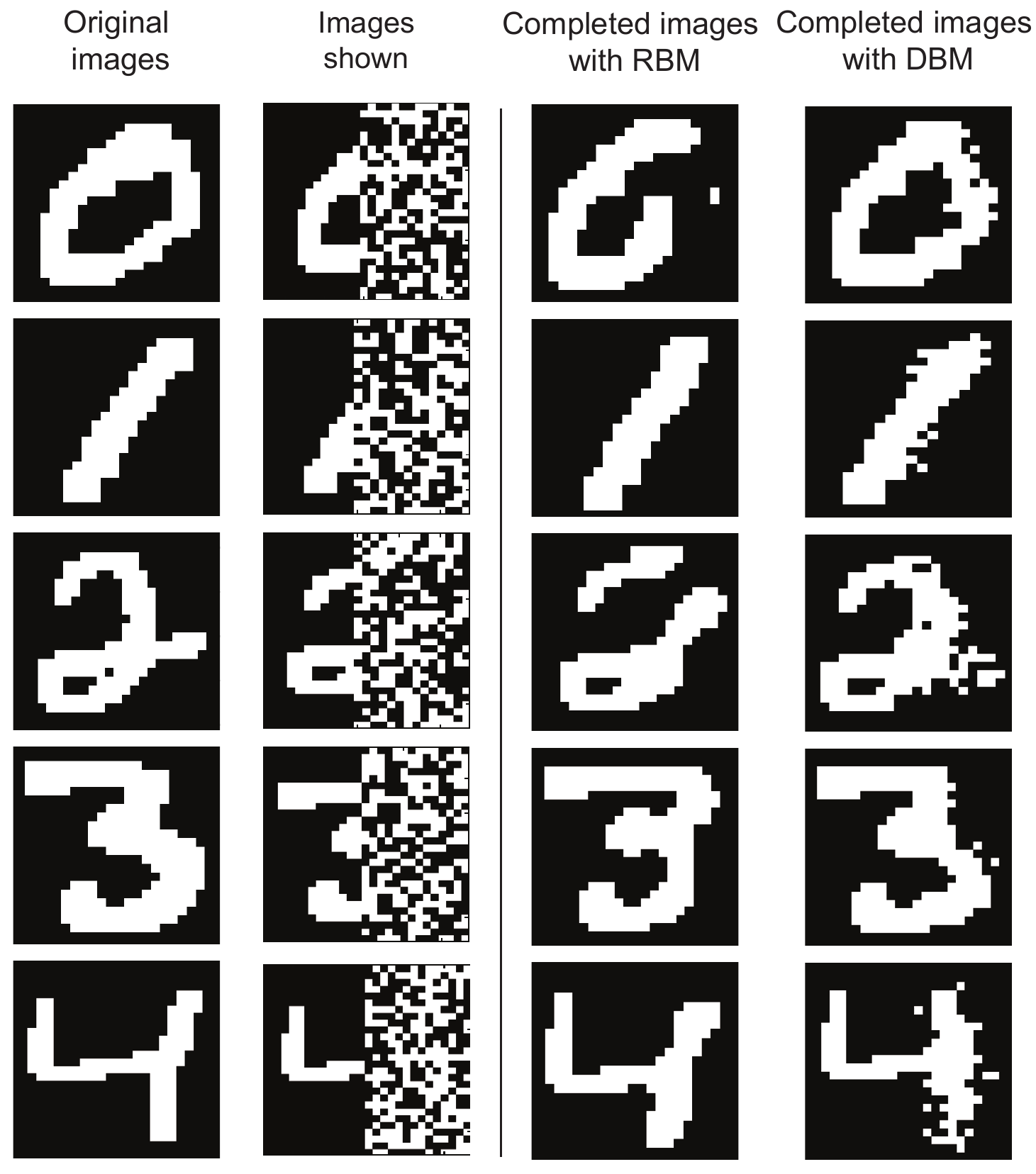}
    \caption{ Image completion examples with RBM (4,096 hidden units and CD-100) and sparse DBM (Pegasus 4,264 p-bits). Only the left half of the images is shown (clamped) to the networks while the other half is obscured. The label bits are also clamped and the annealing schedule varies from $\beta$\,=\,0 to  $\beta$\,=\,5 with 0.125 steps.}
    \label{fig:im_completn} 
\end{figure*}

A typical task for energy-based generative models is that of image completion as opposed to image generation. Image completion is relatively easier since the network is clamped near local minima. In the main manuscript, we show how an iso-accuracy RBM with around 3M parameters cannot perform image generation. Here, we show that an RBM can complete the easier task of image completion.  We clamp half of the visible bits along with the labels and clamp the other half to noise. Then we let the trained network evolve to the corresponding minima by annealing the network from $\beta$\,=\,0 to  $\beta$\,=\,5. Results are shown for a sparse DBM (4,264 p-bits) and RBM (4,096 hidden units, trained with CD-100).  We observe that despite failing at image generation, the RBM performs similarly to our sparse DBM in image completion as shown in FIG.~\ref{fig:im_completn}. 

\vspace{-15pt}
\subsection{Quality of parallel samples}
\label{sec:exactness_samples}

\begin{figure*}[t]
\centering
\includegraphics[width=0.45\linewidth]{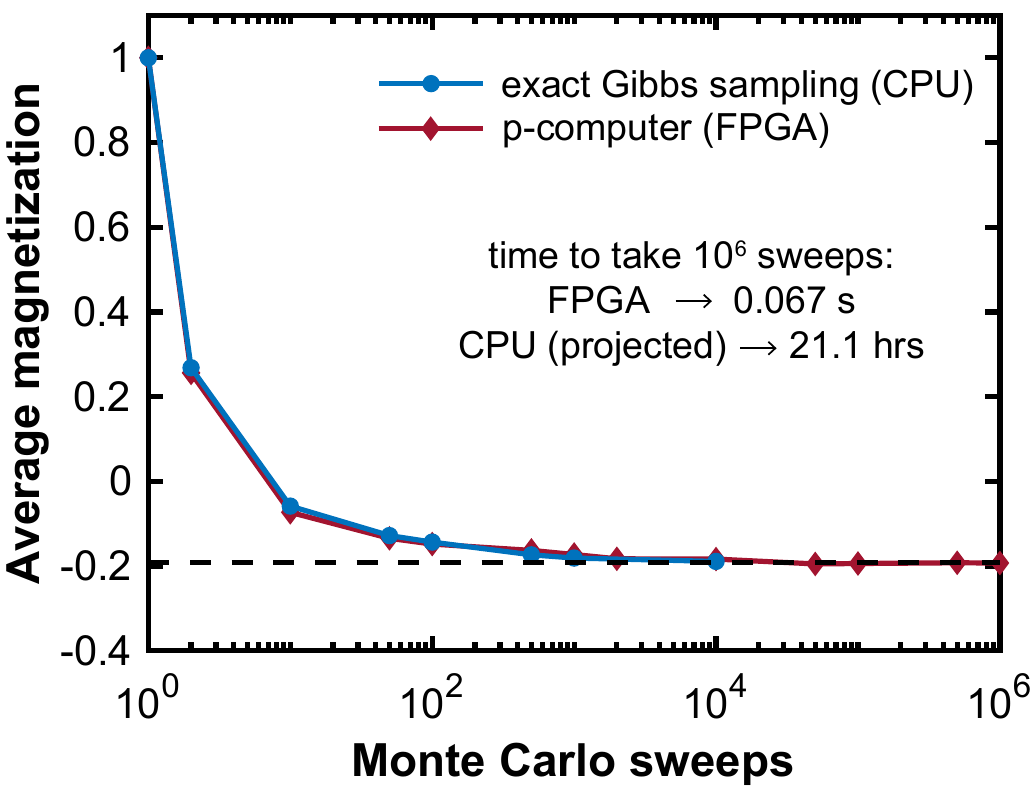}
\vspace{-10pt}
    \caption{Average magnetization as a function of Monte Carlo sweeps in a CPU (exact Gibbs sampling) vs. parallelized FPGA. See text for details.}
    \label{fig:avg_mz}
    \vspace{-15pt}
\end{figure*}

To explicitly show the quality of our parallel samples in our graph-colored architecture (in the main Section~\ref{sec:architecture}), we have performed the following ``inference'' experiment in the CPU performing exact Gibbs sampling vs. our parallelized FPGA using a sparse DBM (Pegasus 3,080 p-bits):\vspace{-5pt}
\begin{itemize}[noitemsep]
\item We start with an MNIST-\emph{trained} Pegasus network with (3080 p-bits) with known weights and biases. 
\item We initialize all p-bits to the +1 state at time step 1 and define a ``network magnetization'', $ \langle m \rangle_k  = \sum_i^N (m_i) /N $
\item We perform Exact (sequential) Gibbs Sampling in a CPU and our parallelized Gibbs Sampling in the FPGA for M=100 times, measuring $\langle m \rangle_k$ for each run, $k \in {1, \ldots, M}$. We obtain an \emph{ensemble-averaged} $\langle m \rangle$.
\item We then compare these averaged magnetizations, \emph{as a function Monte Carlo sweeps} taken in the FPGA and CPU. 
\end{itemize}\vspace{-5pt}
\noindent FIG.~\ref{fig:avg_mz} shows the results of this experiment showing near identical relaxation between the FPGA and the CPU. FPGA takes about 0.067 seconds to take a million samples as opposed to a projected 21.1 hours from the CPU (we did not take more than 10,000 samples over 100 iterations in the CPU, since at that point both models converged). These numbers are in accordance with our expectations from their relative flips/second numbers and they establish that the samples taken by the FPGA follow the Gibbs sampling process. 

\subsection{Grayscale training of full Fashion MNIST}
\label{sec:fashionMnist}
 To test the differences between sparse DBMs and RBMs, we used our largest network (Pegasus 4264 p-bits) to train full Fashion MNIST \cite{xiao2017fashion}, a more challenging dataset than MNIST \cite{liao2022gaussian}. Fashion MNIST consists of 28$\times$28 grayscale images of 70,000 fashion products (60,000 in the training set and 10,000 in the test set) from 10 categories (e.g. t-shirt/top, trouser, sneaker, bag, pullover, and others), with 7,000 images per category. We have trained this dataset using our sparse DBM on the Pegasus graph. There are  4264 p-bits with 30,404 parameters in this network where the number of visible units is 784, 50 label units, and 3430 hidden units.
 
 Our approach to grayscale images is based on time-averaging inspired by the stochastic computing approach of Ref.~\cite{hirtzlin2019stochastic}, where grayscale images between 0 and 1 are treated as the time-averaged probability of activation for p-bits. During the positive phase, we choose N (e.g., 20, 50, 100) binary samples from this probability and clamp the visible nodes as described in the main Section~\ref{sec:algorithm} and Section~\ref{sec:train}. Here, we used 1200 mini-batches containing 50 grayscale images in each batch during training. To train  Fashion MNIST, we used 20 binary (black and white) samples for each grayscale image resulting in a total of 60,000$\times$20 training images. We found that the number of black and white samples can vary depending on the dataset as a hyperparameter.

\begin{table*}[h]
    \centering
    \caption{ Fashion MNIST accuracy with different sizes of RBMs.}
    \vspace{4pt}
    \begin{tabular}{@{}cccc@{}}
        \toprule
        {\bf Number of} & {\bf number of } & {\bf maximum} \\ 
        {\bf hidden units} & {\bf  parameters} & {\bf accuracy (\%)} \\ \midrule
        43 & $34\times 10^{3}$  & 71.90 \\
        64 & $50\times 10^{3}$  & 76.56 \\
        128 & $101\times 10^{3}$  & 77.45\\
        256 & $203\times 10^{3}$  & 78.45\\
        1264 & $1\times 10^{6}$  & 85.56\\
        2048 & $2\times 10^{6}$  & 84.72\\
        4096 & $3.25\times 10^{6}$  & 82.64\\
        \bottomrule
         \end{tabular}
   \label{tab:rbm_fmnist}
\end{table*}

\begin{figure*}[h]
\centering
\includegraphics[width=0.80\linewidth]{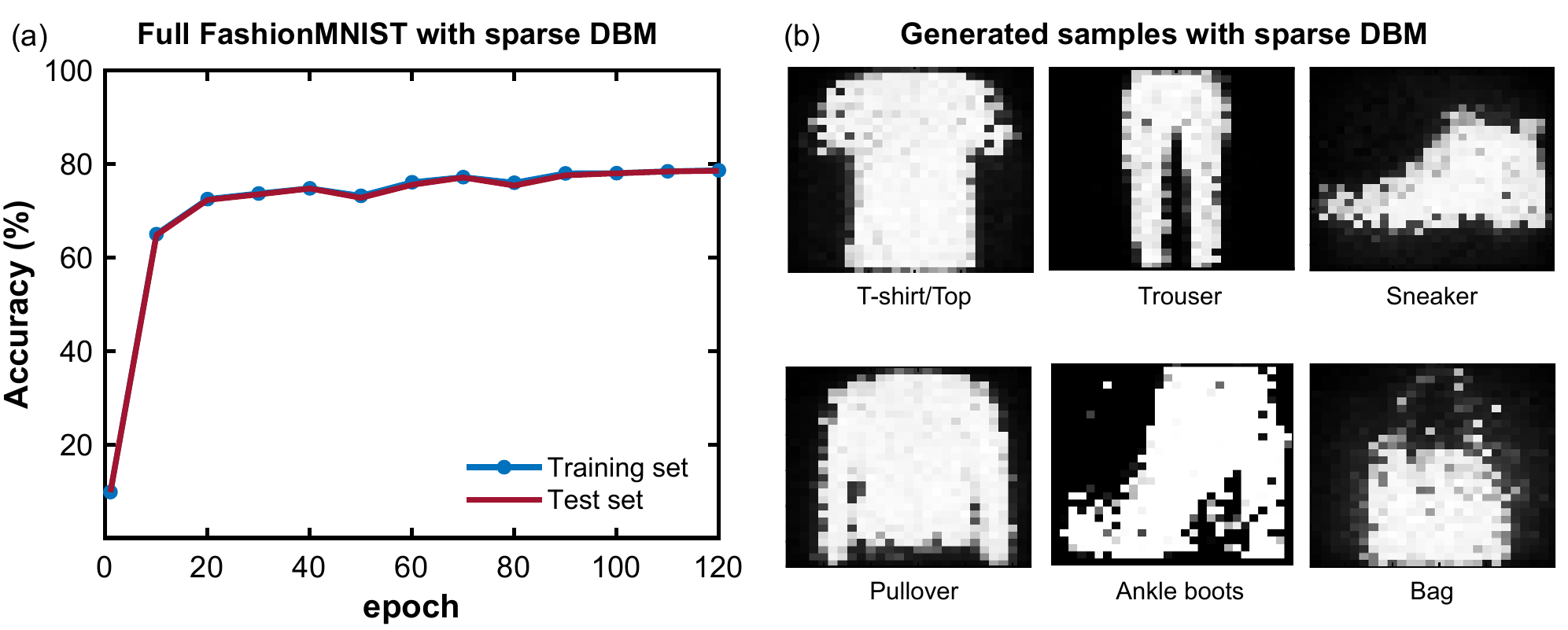}
\vspace{-10pt}
    \caption{(a) Fashion MNIST accuracy can reach around 80\% in 120 epochs with sparse DBM. Full Fashion MNIST (60,000 images) with 20 black and white samples per grayscale image is trained on sparse DBM (Pegasus 4,264 p-bits and 30,404 parameters) using CD-$10^{5}$, batch size = 50, learning rate = 0.003, momentum = 0.6 and epoch = 120. Test accuracy shows the accuracy of the whole test set (10,000 images) and the training accuracy represents the accuracy of 10,000 images randomly chosen from the training set. (b) Image generation examples with sparse DBM after training the network with full Fashion MNIST dataset by annealing the network from  $\beta$\,=\,0 to  $\beta$\,=\,1 with 0.125 steps.}
    \label{fig:fmnist}
    \vspace{-15pt}
\end{figure*}

 To test classification accuracy, we also used 20 black and white samples for each grayscale image to perform the same softmax classification as described in the main Section~\ref{sec:train}. After the network sees those 20 black and white samples to form a grayscale image, we check the labels to establish classification accuracy. Using this scheme of grayscale images, our sparse DBM with 30,404 parameters can reach around 80\% in 120 epochs as shown in FIG.~\ref{fig:fmnist}a.
 We trained RBMs with different numbers of parameters as listed in Table~\ref{tab:rbm_fmnist} using the same approach as sparse DBM. The iso-parameter RBM (43 hidden units and 34k parameters) can reach a maximum of 72\% accuracy on full Fashion MNIST while the million-parameter RBMs can go to around 85\% test accuracy. As in MNIST, we see that RBM requires the order of a million parameters to reach sparse DBM accuracy.

 Using a similar approach to image generation as described in the main Section~\ref{sec:ImageSynth}, we can also generate images of fashion products with our sparse DBM as shown in FIG.~\ref{fig:fmnist}b. Similar to other cases, for image generation, we only clamp label bits for a particular image. We  anneal the network slowly from $\beta$\,=\,$0$ to $\beta$\,=\,$5$ with a $0.125$ increment using the final weights and biases. Then we check the 784 visible p-bits after time-averaging the collected samples to obtain grayscale images. Using a similar procedure, we observe that none of the RBMs can generate images despite having a maximum of 85\% accuracy. We observed that different annealing schedules (e.g., with slower $\beta$ changes)  do not help RBM image generation.

\subsection{Grayscale CIFAR-10/100}
\label{sec:cifar100}

To see whether the same conclusions hold for our architectural and algorithmic ideas for sparse DBMs,  we trained 100 images (10 images from each class) from the CIFAR-10 dataset \cite{CIFAR-10and100}. Due to resource limitations in our FPGA, we could not increase the number of parameters, hence we used Pegasus 4,264 p-bits as sparse DBM to train the grayscale CIFAR-10/100. The CIFAR-10 dataset consists of 32$\times$32 color images of 10 different classes (i.e., airplane, automobile, bird, cat, deer, dog, frog, horse, ship, and truck), with 6000 images per class. There are 50000 training images and 10000 test images.

\begin{figure*}[!h]
\centering
\includegraphics[width=0.95\linewidth]{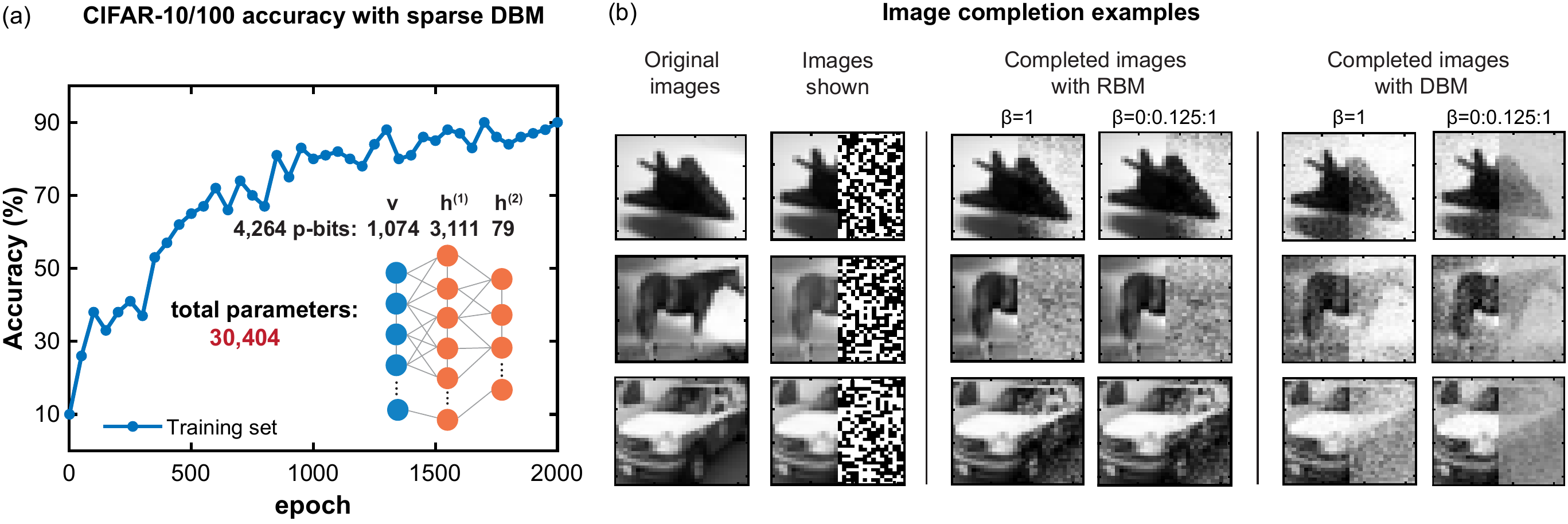}
\vspace{-10pt}
    \caption{(a) Training accuracy of 100 images from CIFAR-10 is around 90\% in 2000 epochs with sparse DBM. Training is accomplished with 100 black and white samples per grayscale image using CD-$10^{5}$, batch size = 10, learning rate = 0.006, momentum = 0.6, and epoch = 2000. (b) Image completion examples with RBM (4096 hidden units) and sparse DBM. Only the left half of the grayscale images are shown (clamped) to the networks (using 100 black and white samples) while the other half is obscured. The label bits are also clamped and the annealing schedule varies from $\beta$\,=\,0 to  $\beta$\,=\,5 with 0.125 steps, and for the other case $\beta$ is kept to 1.}
    \label{fig:cifar_accuracy}
    \vspace{0pt}
\end{figure*}

We converted the color images into grayscale \cite{qin2021optimizing} using the `rgb2gray' function in MATLAB. We used 1024 visible units, 5 sets of labels (50 p-bits), and 3190 hidden units arranged in 2 layers as shown in the inset of FIG.~\ref{fig:cifar_accuracy}a. Utilizing the same approach of binary transformation from grayscale images as described in Section~\ref{sec:fashionMnist}, we have trained the CIFAR-10/100 dataset with 100 black and white samples per grayscale image with 10 mini-batches (having randomly selected 10 grayscale images in each batch). Training accuracy of this dataset with sparse DBM can reach around 90\% in 2000 epochs as shown in FIG.~\ref{fig:cifar_accuracy}a while the iso-parameter RBM (40 hidden units) accuracy is only 68\% as listed in Table~\ref{tab:rbm_cifar100}. RBMs reach the same levels of accuracy using between 264,000 and 1 million parameters, in line with our earlier results.

Image generation for CIFAR-10 in this reduced setting with only 100 images in the training set failed both for sparse DBM and RBM. For this reason,  we also examined the image completion task (details in Section~\ref{sec:im_compltn_dbm_rbm}) with RBM and sparse DBM as shown in FIG.~\ref{fig:cifar_accuracy}b. We clamped only the left half of a grayscale image (using 100 black and white samples) along with the corresponding label bits and checked the right half of that image. In this case, both RBM (4096 hidden units) and sparse DBM performed similarly, in this much harder setting.

\begin{table*}[h]
    \centering
     \caption{{ CIFAR-10/100 accuracy with different sizes of RBMs.}}
     \vspace{4pt}
    \begin{tabular}{@{}cccc@{}}
        \toprule
        {\bf Number of} & {\bf number of } & {\bf maximum} \\ 
        {\bf hidden units} & {\bf  parameters} & {\bf accuracy (\%)} \\ \midrule
        40 & $41\times 10^{3}$  & 68 \\
        64 & $66\times 10^{3}$ & 83 \\
        128 & $132\times 10^{3}$  & 88\\
        256 & $264\times 10^{3}$  & 88\\
        968 & $1\times 10^{6}$  & 99\\
        2048 & $2\times 10^{6}$   & 100\\
        4096 & $4\times 10^{6}$   & 100\\
        \bottomrule
         \end{tabular}
   \label{tab:rbm_cifar100}
\end{table*}

\subsection{Full graph topologies: Pegasus 4,264}
\label{sec:actual}
Below we show the full Pegasus network topology with 4,264 p-bits and its sparse deep BM representation. 

\begin{figure*}[!h]
    \centering
    \includegraphics[width=0.68\textwidth]{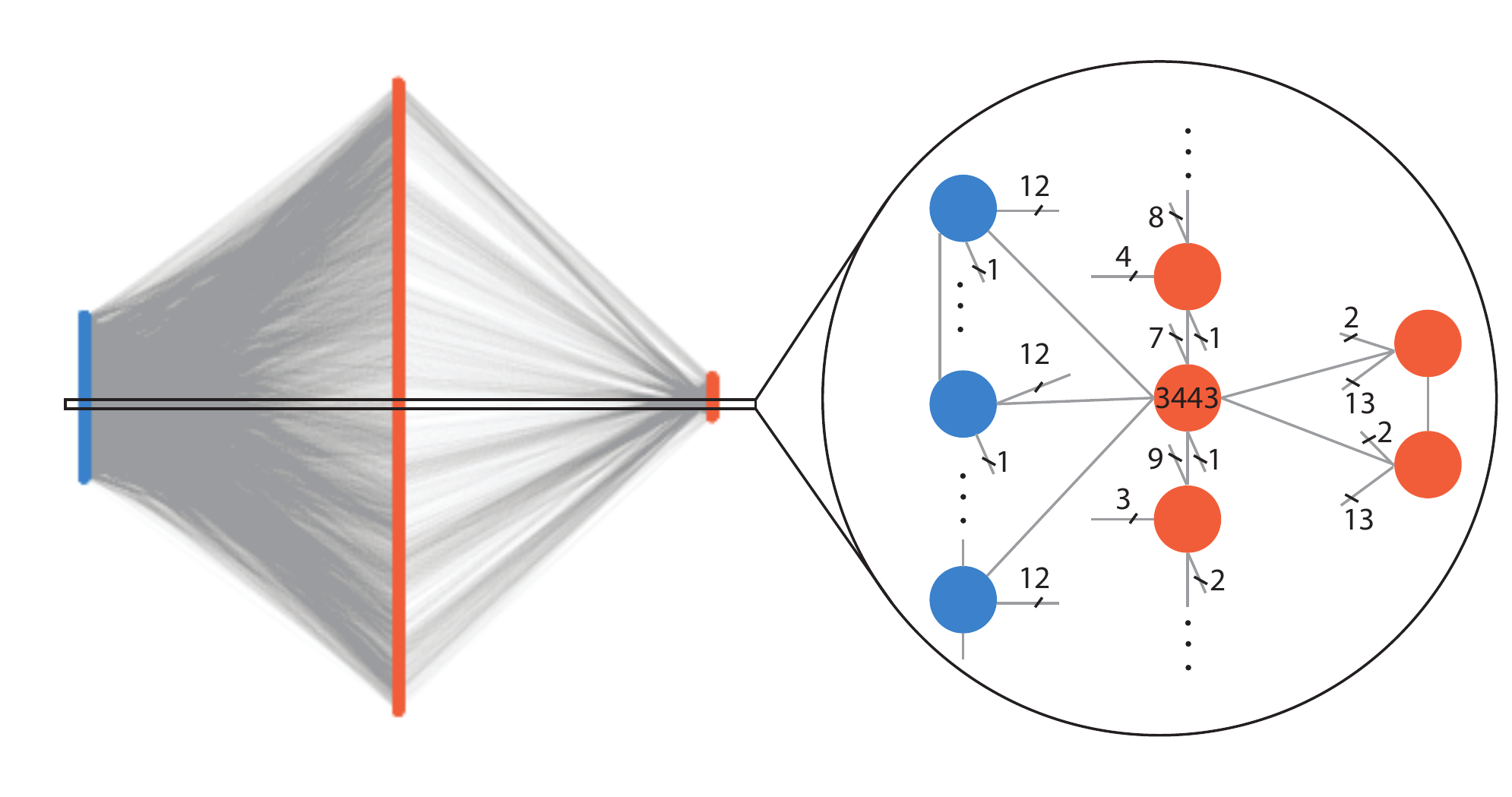}
    \caption {{\footnotesize Layered embedding of the 4,264 p-bit Pegasus graph of FIG.~\ref{fig:P14}, illustrating the sparse DBM architecture: the first layer is visible p-bits with 834 nodes, second and third layers are the hidden p-bits with 3,226 and 204 nodes respectively. There are also some intralayer connections within each layer. An example is shown in the right circle which shows the neighboring connections around node 3,443. The number next to a line represents the number of wires grouped in that branch, the total number being the fan-out of a given p-bit (vertex).}}
\label{fig:P14_layered}
\end{figure*}

\begin{figure*}[h]
    \centering
    \includegraphics[width =0.95\textwidth ]{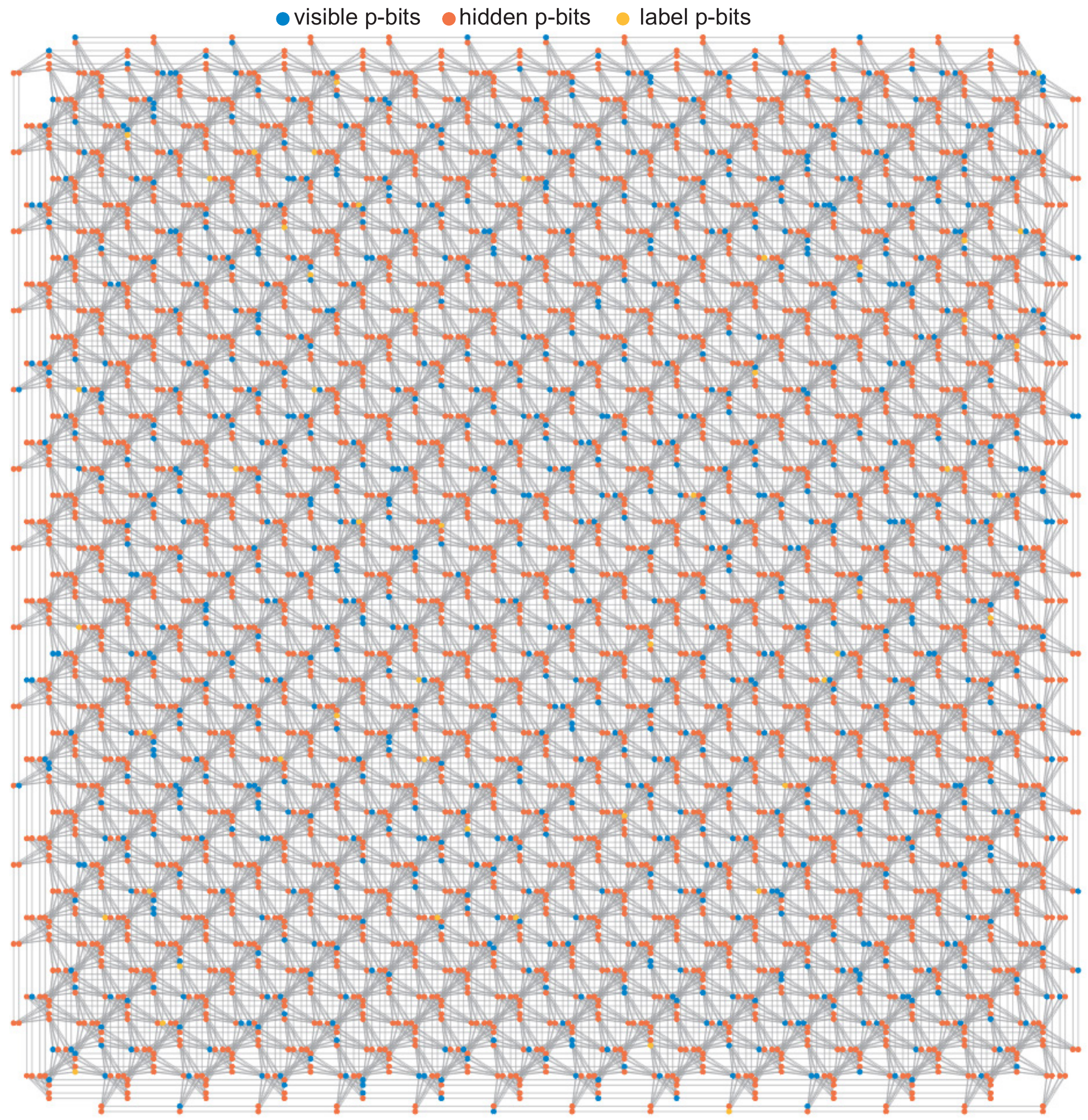}
    \caption {{\footnotesize The original sparse DBM network (Pegasus: 4,264 p-bits) used in this work with marked-up visible (blue), hidden (orange), and label (yellow) units.}}
\label{fig:P14}
\end{figure*}

\end{document}